\documentclass[trackchanges,twocolumn,twocolappendix]{aastex701}
\usepackage{amsmath}
\usepackage[normalem]{ulem}

\definecolor{comm}{rgb}{0,0.7,0}

\newcommand{\code}[1]{\texttt{#1}}

\newcommand{\mesa}{\code{MESA} }
\newcommand{\MESA}{\mesa}

\newcommand{\gyre}{\code{GYRE} }
\newcommand{\GYRE}{\gyre}

\begin{document}

\title{Solar-like Oscillations in Accreting Pre-Main Sequence Stars: Insights and Prospects}

\author{Johannes Jørgensen}
\affiliation{Universit\"at Innsbruck, Institut f\"ur Astro- und Teilchenphysik, Technikerstra{\ss}e 25, 6020 Innsbruck, Austria}
\email[show]{johannes.joergensen@uibk.ac.at} 

\author{Konstanze Zwintz}
\affiliation{Universit\"at Innsbruck, Institut f\"ur Astro- und Teilchenphysik, Technikerstra{\ss}e 25, 6020 Innsbruck, Austria}
\email[]{Konstanze.zwintz@uibk.ac.at}

\author{Ebraheem Farag}
\affiliation{Department of Astronomy, Yale University, New Haven, CT 06511, USA}
\email[]{ebraheem.farag@yale.edu}

\author{Eduard I. Vorobyov}
\affiliation{University of Vienna, Department of Astrophysics, Türkenschanzstrasse 17, 1180 Vienna, Austria}
\email[]{}

\author{Thomas Steindl}
\affiliation{Universit\"at Innsbruck, Institut f\"ur Astro- und Teilchenphysik, Technikerstra{\ss}e 25, 6020 Innsbruck, Austria}
\email[]{Thomas.Steindl95@gmx.at}

\begin{abstract}
We present theoretical predictions for solar-like oscillators in the pre-main sequence phase of stellar evolution. Our pre-main sequence models start from a stellar seed of 0.01 solar masses that gains mass through accretion, offering an alternative description to the classical approach segmented into the Hayashi and Henyey tracks. Evolutionary models are calculated using the \mesa stellar evolution code with a custom accretion routine and pulsation properties are investigated using the \GYRE oscillation code. We present evolutionary tracks and internal structures for accreting pre-main sequence solar-like stars in the mass range from 0.7 to 1.6 solar masses, adopting 35 mass accretion histories previously extracted from two-dimensional magneto-hydrodynamical simulations. Atmospheric parameters of our models constrain characteristic frequencies of pre-main sequence solar-like oscillators to be generally greater than 500 $\mu\text{Hz}$. We highlight the imprint of accretion on the buoyancy and Lamb profiles and illustrate the effects on the small- and large frequency separations. We additionally quantify individual frequency differences across the 35 accretion histories at the zero-age main-sequence, showcasing differences no larger than $20 \ \mu\text{Hz}$ for an exemplary model. Finally, we discuss the potential of detecting solar-like oscillations in pre-main sequence stars with the upcoming ESA PLATO mission.
\end{abstract}



\section{Introduction}

Asteroseismology utilizes surface pulsations of stars to probe their interiors and constrain the input physics of stellar models \citep{Aerts(2021):Probing_the_interior_of_stars_with_asteroseismology}. The pristine photometric data from space-based missions such as NASA's Kepler \citep{Boruck(2010):Kepler} and Transiting Exoplanet Survey Satellite (TESS) \citep{Ricker(2014):TESS} have especially allowed for the study of stellar interiors through detected pulsations. The future ESA (European Space Agency) space mission PLATO (PLAnetary Transits and Oscillations of stars, expected launch in December 2026) is expected to observe up to $1 \ 000 \ 000$ stars and will similarly further advance our ability to  describe the structure and evolution of stars and their planetary systems \citep{Rauer(2025)}. Despite these increasing efforts, pulsations occurring during the pre-main sequence (pre-MS) phase of stellar evolution remain challenging to characterize due to limited observational data and the reliance on less computationally demanding models of early stellar evolution \citep{Zwintz&Steindl(2022)}. Joint studies of asteroseismology and realistic scenarios of early stellar evolution are, for example, carried out in \cite{Steindl(2021)I}, \cite{Steindl(2022)II} and \cite{Steindl(2022):Impact_of_accretion} where theoretical models of pre-MS $\delta$-Scuti stars, for which pulsations are driven by the $\kappa$-mechanism, that have undergone mass accretion to attain their zero-age main-sequence (ZAMS) masses are calculated. The authors establish pre-MS instability strips, perform asteroseismic modeling of the pre-MS $\delta$-Scuti star HD 139614, and characterize the pulsation properties of the pre-MS models. Furthermore, while not directly concerning pre-MS stars, authors such as \cite{Serenelli(2011)} or \cite{Kunitomo(2021)} have also proposed protoplanetary disk accretion in solar models as a solution to the so-called solar abundance problem \citep{Asplund(2006)} with the solar interior inferred from helioseismology \citep{JCD(1996):helioseismology, Basu(2016)}.

In the work presented here, we extend the pre-MS studies to lower masses and a different excitation mechanism, that is, to the theoretically predicted class of pre-MS solar-like oscillators (\citealp{Samadi(2005)}, \citealp{Pinheiro(2008)}). In these objects turbulent envelope convection is expected to drive surface pulsations as observed in the Sun, low-mass main-sequence stars, and red giants \citep{JCD(2004):Physics_of_SLO, Garcia(2019)}. Apart from a single candidate detection \citep{Mullner(2021)}, and an exploration on rotational dampening of oscillations in pre-MS stars \citep{Bessila(2024)}, this class of objects remains mostly unexplored, yet. 

We present low-mass pre-MS stellar models undergoing mass accretion. These are compared to models calculated from the classical scenario where early stellar evolution is segmented into Hayashi and Henyey tracks \citep{Henyey(1955), Hayashi(1961), Iben(1965)}. We explore the parameter space of our models and discuss their corresponding pulsation properties. Finally we put our findings in the context of the PLATO mission.

\section{Methods}
We first outline our \mesa and \GYRE setup for evolving accreting protostars and calculating their pulsation frequencies.

\subsection{Stellar models with MESA}

We calculate stellar models with \MESA r22.11.1 \citep{Paxton2011, Paxton2013, Paxton2015, Paxton2018, Paxton2019, Jermyn2023}. Convection is treated according to \cite{Henyey(convection)} with the Ledoux criteria for convective instability \citep{Ledoux(1947)} and $\alpha_{\text{mlt}}=1.7$ for the mixing length parameter. Exponential convective overshooting of \cite{Herwig(2000)} is also turned on with over- and undershooting parameters $f/f_0=0.01/0.005$ and $f/f_0=0.005/0.0025$ respectively. We additionally include semiconvection from \cite{Langer(1983)} with $\alpha_{\text{sc}}=0.1$ and thermohaline mixing following \cite{Kippenhahn(thermohaline)} with $\alpha_{\text{th}}=17.5$ \citep{Posydon}. We impose a minimum value of \(20\,\mathrm{cm}^2\,\mathrm{s}^{-1}\) for the diffusive mixing coefficient \citep{Wagg(2024)}. For initial parameters we choose a solar chemical composition with $Y=0.275$ and $Z=0.018$ (see for example \citealp{Bellinger(2022)}) and masses in the range 0.7 to 1.6 M$_\odot$. The abundances of $^2\text{H}$ and $^3\text{He}$ are additionally set to 20 ppm and 85 ppm. Our \MESA inlists are made publicly available on Zenodo\footnote{\url{https://doi.org/10.5281/zenodo.17424743}} and we refer to the appendix for a discussion on numerical convergence. The scheme for implementation of accretion luminosity in \MESA is adopted from \cite{Steindl(2022):Impact_of_accretion} which in addition is based on previous work from \cite{Siess(1997)}, \cite{Baraffe(2009)} \cite{Kunitomo(2017)}, \cite{SJensen(2018)} and \cite{Elbakyan(2019)}.

We stress that our results hinge on the physical interpretation of our evolutionary models. While realistic accretion histories are used, not all physical processes known to be present during star formation are included. Magnetic fields and rapid rotation, for example, influence the evolution of young stars, but incorporating these effects is beyond the scope of this work \citep{Zwintz(2009), Alencar(2010), Cody(2014), Zwintz(2019), Zwintz&Steindl(2022)}. Pulsation frequencies are likewise affected by these phenomena but not accounted for \citep{Chaplin(2000):Magnetic_dampening, Jenkins(2011), Garcia(2011), Bonanno(2014), Bessila(2024)}. The initial abundances of $^2\text{H}$ and $^3\text{He}$ will also affect evolutionary tracks; however, for simplicity, these parameters were fixed. The treatment of accretion outlined in Sec. \ref{subsec:modelling_accretion} additionally relies on simplified assumptions regarding the deposition of accretion energy. The exact nature is not fully understood and results may vary depending on how it is treated (see, for example, \citealp{Baraffe(2009), Kunitomo(2017), SJensen(2018), Elbakyan(2019), Steindl(2021)I, Steindl(2022)II, Steindl(2022):Impact_of_accretion, Zwintz&Steindl(2022)}).

\subsubsection{Initial stellar seed}
The \MESA calculations begin from an initial protostellar seed of 0.01 M$_\odot$ and 1.5 R$_\odot$ corresponding to a second Larson core with significant entropy \citep{Larson(1969)}. In general, turbulence within the cloud leads to the formation of gravitationally bound condensations that will collapse under their own gravity. As the contracting condensation becomes opaque to its own radiation, the first hydrostatic core forms in its depths. Subsequent accretion on the opaque core raises temperatures above 2000 K at which point molecular hydrogen is fully dissociated and the protostellar seed is formed. An initial \MESA run generates this seed by first creating a 0.03 M$_\odot$ pre-MS model with same initial parameters as in the subsequent accreting evolution. Using the standard mass-relax scheme in \MESA 0.02 M$_\odot$ are removed and the remaining model is allowed to contract until it reaches 1.5 R$_\odot$. Protostellar seeds with similar characteristics can form in the gravitational collapse of an interstellar gas cloud \citep{Masunaga(2000)}. The seed subsequently accretes matter until the final ZAMS mass is reached. 

We note here that the choice to begin the accretion phase from initial models of $0.01 \ \text{M}_\odot$ and $1.5 \ \text{R}_\odot$, mimicking protostellar seeds of significant entropy, is motivated mainly by convenience of numerical stability \citep{Kunitomo(2017)}. However, while the subsequent evolution is mostly indifferent to variations in seed radii \citep{Baraffe(2009)}, this is not the case for variations in seed masses. Previous studies, for example \cite{Baraffe(2012)}, examined seed masses of 1-5 Jupiter masses and found generally hotter and more luminous objects emerging from the accretion phase. Another study, \cite{Haemmerle(2019)}, opted to skip the seed phase entirely, initializing their models at a minimum of 0.7 M$_\odot$. In short, our results are consistent only in the framework of initiating the accretion phase from a seed of $0.01 \ \text{M}_\odot$ but can withstand variations in the seed radius by a factor of several.

\subsubsection{Modeling accretion}\label{subsec:modelling_accretion}
We implement accretion in \MESA as a non-spherical, episodic process that allows the star to radiate away its energy over most of the photosphere. We first consider the kinetic energy of the infalling material,
\begin{equation}\label{eq:ekin}
    E_\text{kin} = \frac{1}{2}Mv^2,
\end{equation}
where $M$ is the mass of the infalling material. From conservation of energy we write the infall velocity as,
\begin{equation}\label{eq:vel}
    v=\sqrt{\frac{2GM_\star}{R_\star}},
\end{equation}
where $M_\star$ and $R_\star$ are the mass and the radius of the star. Combining Eqs. \ref{eq:ekin} and \ref{eq:vel} and calculating the derivative with respect to time we obtain the rate of kinetic energy of the infalling material:
\begin{equation}
    \frac{\text{d}E_\text{kin}}{\text{d}t} = G\frac{\dot{M}M_\star}{R_\star},
\end{equation}
where $\dot{M}$ is rate of change of the infalling material, that is, the mass accretion rate. We now introduce a parameter ($\epsilon\leq1$) that represents the geometry of the system and the fraction of the total energy that is actually transferred into heat and radiated away,
\begin{equation}
    L_\text{acc} = \epsilon G\frac{\dot{M}M_\star}{R_\star}.
\end{equation}
In the thin-disk limit we have $\epsilon=0.5$ meaning half the energy is radiated away. Lastly, we introduce a second parameter ($\beta\leq1$) which controls the thermal efficiency of the accretion process. We thus get the energy injected into the star,
\begin{equation}
    L_\text{add} = \beta\epsilon GM\dot{M}/R,
\end{equation} 
and the energy radiated away,
\begin{equation}\label{eq:lacc}
    L_\text{acc} = (1-\beta)\epsilon GM\dot{M}/R.    
\end{equation}
The limit $\beta=0$ is referred to as cold accretion, where no energy is absorbed, and $0<\beta\leq1$ is denoted hot accretion. We parametrize $\beta$ by $\dot{M}$ as a step function with a smooth transition:
\begin{equation}\label{eq:beta}
    \beta(\dot{M}) = \frac{\beta_\text{L} \ \text{exp}\left(\frac{\dot{M}_\text{m}}{\Delta}\right) + \beta_\text{U} \ \text{exp}\left(\frac{\dot{M}}{\Delta}\right)}{\text{exp}\left(\frac{\dot{M}_\text{m}}{\Delta}\right) + \text{exp}\left(\frac{\dot{M}}{\Delta}\right)},
\end{equation}
where $\beta_\text{L} = 0.005$ and $\beta_\text{U} = 0.2$ are the lower- and upper bounds of $\beta$ and $\dot{M}_\text{m} = 6.2\times10^{-6} \ \text{M}_\odot \ \text{yr}^{-1}$ and $\Delta = 5.95\times10^{-6} \ \text{M}_\odot \ \text{yr}^{-1}$ are the midpoint and the width of the crossover between the lower and upper limits \citep{SJensen(2018), Steindl(2022):Impact_of_accretion}. The energy of accreted matter $L_\text{add}$ is added as heat which is distributed through a linear increase as a function of mass:
\begin{equation}
    l = \frac{L_{\text{add}}}{M_\star} \text{max}\left\{0, \frac{2}{M^2_{\text{outer}}}\left(\frac{m_r}{M_\star}-1+M^2_{\text{outer}}\right)\right\},
\end{equation}
where $M_{\text{outer}}$ defines the fractional mass in which the extra heat $l$ is deposited and $m_r$ is the mass coordinate. We parametrize also $M_{\text{outer}}$ with $\dot{M}$ as a step function with a smooth transition:
\begin{equation}\label{eq:Mouter}
    M_{\text{outer}}(\dot{M}) = \frac{q_\text{L} \ \text{exp}\left(\frac{m}{w}\right) + q_\text{U} \ \text{exp}\left(\frac{\dot{M}}{w}\right)}{\text{exp}\left(\frac{m}{w}\right) + \text{exp}\left(\frac{\dot{M}}{w}\right)},
\end{equation}
where here $q_\text{L} = 0.0025$ and $q_\text{U} = 0.2$ are the lower- and upper bounds of $M_{\text{outer}}$ \citep{Steindl(2021)I} and $\log_{10}{m} = -6$ and $\log_{10}{w} = 0.5$ are the midpoint and the width of the crossover between the lower and upper limits in units M$_\odot \ \text{yr}^{-1}$. The values for parameters in Eqs. \ref{eq:beta} and \ref{eq:Mouter} are chosen so that we cover the lower and the upper limits of $\beta$ and $M_\text{outer}$ given the minimum and maximum values of mass accretion rates we have available, and to ensure a smooth transition between the lower and upper limits. In Figure \ref{fig:app:beta_mouter} (App. \ref{app:sec:supplementary_figures}) we illustrate both $\beta$ and $M_\text{outer}$ against $\dot{M}$. Given $\dot{M}$ a \MESA routine injects $L_{\text{add}}$ into the model.

\subsubsection{Mass accretion rates}\label{subsec:mass_accretion_rates}
\begin{figure}
    \centering
    \includegraphics[width=1.0\linewidth]{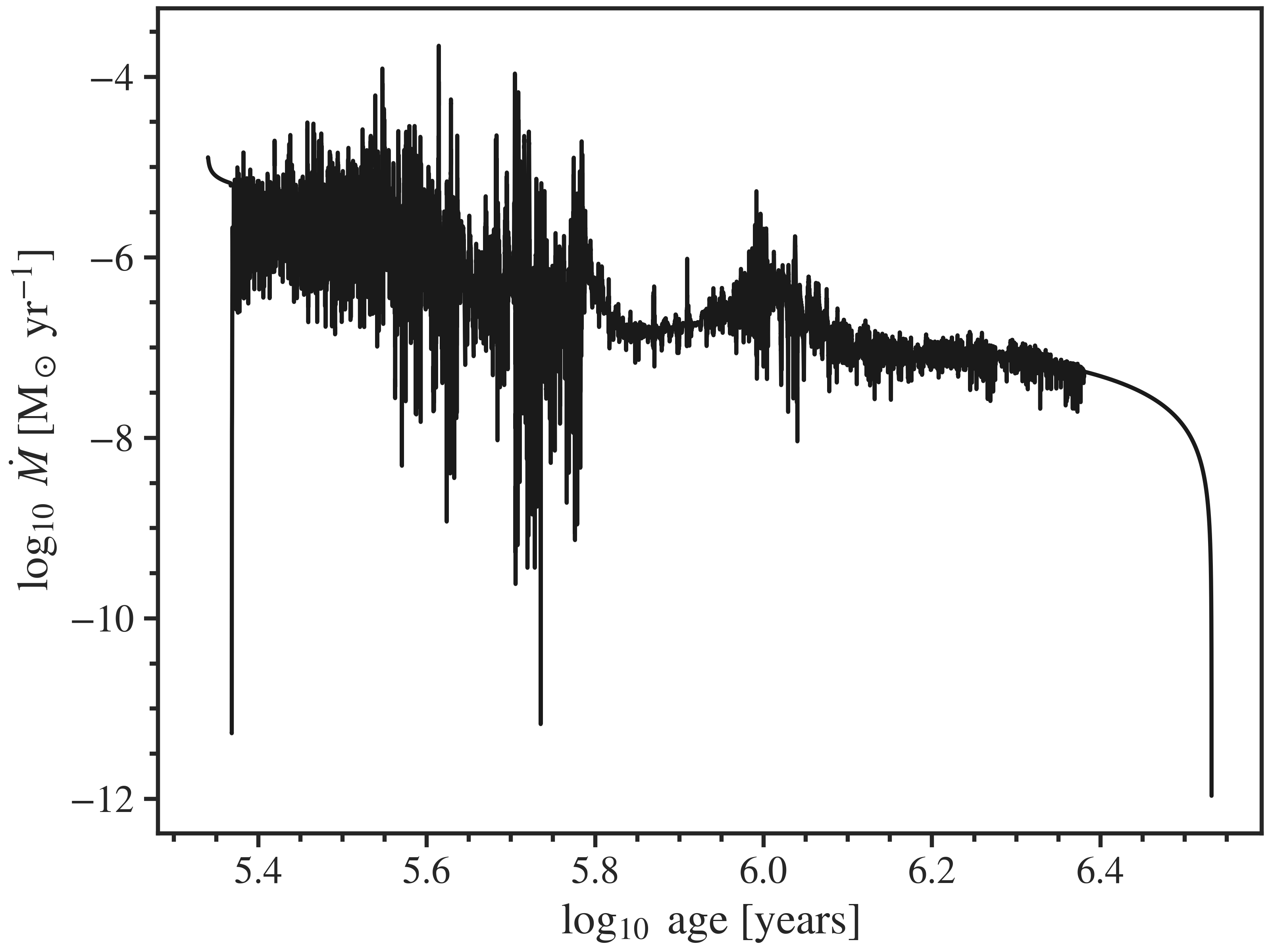}
    \caption{Example accretion history number 35 adopted from \protect\cite{Elbakyan(2019)}.}
    \label{fig:acchist16}
\end{figure}

Mass accretion rates are taken from \cite{Elbakyan(2019)} based on the numerical hydrodynamics model of \cite{Vorobyov(2015)}. This model solves the equations of mass, momentum and energy transport in the thin-disk limit. It furthermore takes into account disc self-gravity, disc surface-cooling due to dust radiation, disc heating via stellar and background irradiation and turbulent viscosity. In contrast to the simplified scenario of a constant accretion rate \citep{Stahler(1980), Palla(1991)}, these mass accretion rates capture the variable nature of star formation, including episodic accretion bursts as manifested by FU Orionis-type eruptions \citep{Fischer(2023), Audard(2024)}. An example accretion history is shown in Figure \ref{fig:acchist16}.

35 accretion histories are available leading to ZAMS masses between 0.0468 and 1.319 $\text{M}_\odot$. By scaling the accretion rates we are able to utilize all 35 accretion histories and cover our desired mass range of 0.7 to 1.6 $\text{M}_\odot$. The scaling procedure is applicable because variable accretion histories with bursts are also typical for stellar objects more massive \citep{Meyer(2017), Vorobyov(2025)} than the ones presented in \cite{Elbakyan(2019)}.

\subsection{Frequency calculations with \GYRE}\label{subsec:GYRE}
We calculate oscillation frequencies using the \GYRE stellar oscillation code \citep{GYRE(2013), GYRE(2018), GYRE(2020), GYRE(2023)}. The full $6^{\rm th}$ order dimensionless stellar oscillation equations \citep{Dziembowski(1971), Christensen_Dalsgaard_2008} are solved with the Colloc scheme \texttt{MAGNUS\_GL6}. We tested grid resolutions of 1000 and 3000 and found no significant differences, hence we settled on 1000. Frequencies are calculated for spherical degrees $\ell=0,1,2,3$ with $\ell=3$ being the maximum degree to which we can resolve surface pulsations with integrated light from space-based photometry. The frequencies are additionally calculated in a frequency range corresponding to $\pm2\cdot0.66\cdot\nu_\text{max}^{0.88}$ (twice that of \citealp{Mosser(2012)}) where $\nu_\text{max}$ is the midpoint of the Gaussian shaped frequency envelope commonly observed in solar-like oscillators. It represents a characteristic frequency and is calculated using the scaling relations of \cite{KjeldsenBedding(1995):Scaling_relations}. The scaling relation for $\nu_\text{max}$ reads
\begin{equation}\label{eq:numax}
    \frac{\nu_\text{max}}{\nu_{\text{max},\odot}} = \frac{(M/M_\odot)}{(R/R_\odot)^2 \sqrt{T_\text{eff}/T_{\text{eff},\odot}}},
\end{equation}
where $M$ is the mass, $R$ is the radius and $T_\text{eff}$ is the effective temperature scaled with the corresponding values for the Sun ($\odot$). An inherent assumption in our work is that the scaling relations hold for pre-MS stars such that we can obtain $\nu_\text{max}$ values for our models \citep{Mullner(2021)}.  We output structure profiles for frequency calculations in tiny increments of luminosity and effective temperature in order to fully map out the evolutionary tracks. Frequencies are additionally calculated in an evolutionary phase when the models are at 70\% of their ZAMS mass up until the end of the MS. For most of the following analysis, however, we present frequencies calculated in a phase when the models have reached 90\% of their ZAMS mass. This is similar to the class II stage of pre-MS stars described in \cite{Vorobyov(2015)} where stars have mostly dissipated their birth environments and become optically visible.

\section{Accreting stellar evolution}
In this section we showcase the evolution of our models and their interiors across the Hertzsprung-Russell (HR) diagram and compare to models calculated using the classical prescription of the pre-MS based on the approach by Hayashi and Henyey. 

\subsection{Evolution in the HR-diagram}

\begin{figure*}
    \centering
    \includegraphics[width=\textwidth]{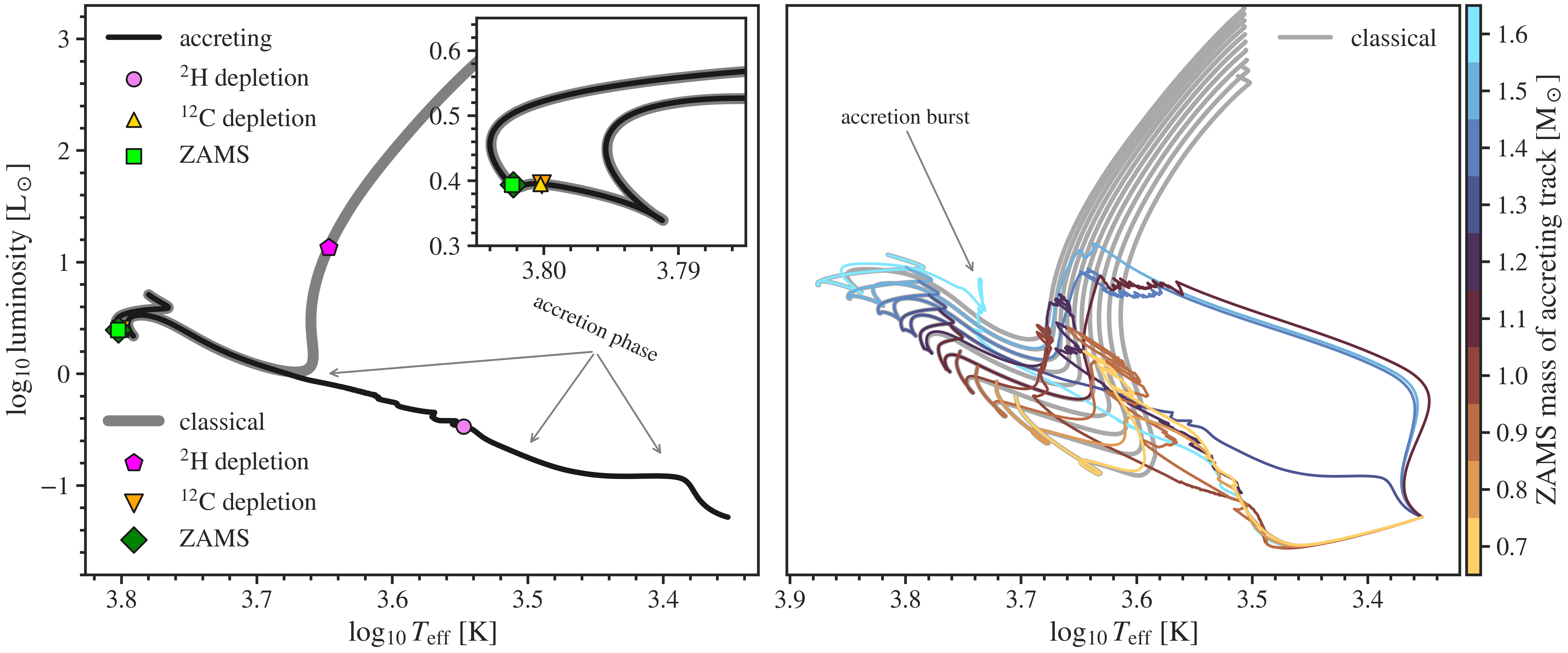}
    \caption{Left: 1.3 M$_\odot$ classical (gray) and accreting evolutionary tracks (black). Only intrinsic luminosity is drawn here, meaning that $L_\text{acc}$ (equation \ref{eq:lacc}) is not included. Symbols highlight key windows during the pre-MS evolution; the depletion of $^2\text{H}$, the depletion of $^{12}\text{C}$ and the ZAMS. The latter two are made visible with an inset figure and the accretion phase is also indicated with arrows. Right: Classical and selected accreting evolutionary tracks in the HR-diagram. Again, only intrinsic luminosities are drawn here, meaning that $L_\text{acc}$ (equation \ref{eq:lacc}) is not included. Evolutionary tracks are drawn until the central hydrogen abundance drops 0.1\% from its initial value. A luminosity burst in the 1.6 M$_\odot$ model is also indicated with an arrow.}
    \label{fig:HR_diagram_big}
\end{figure*}

We first follow the evolution of a single pre-MS model in the HR-diagram. This is illustrated in the left panel of Figure \ref{fig:HR_diagram_big} where we compare 1.3 M$_\odot$ models in the classical (gray) and the accreting scenarios (black). We do not include the accretion luminosity (equation \ref{eq:lacc}) in the total luminosity and therefore only the intrinsic, or photospheric, luminosity is drawn here. The accretion phase is indicated by arrows and the typical tracers of pre-MS evolution are also highlighted. These are, respectively, the depletion of $^2\text{H}$ following the second step in the proton-proton chain, $^2\text{H} \ + \ ^1\text{H} \ \rightarrow \ ^3\text{He} \ + \ \gamma$, the depletion of $^{12}\text{C}$ following the first step in the CNO cycle, $^{12}\text{C} \ + \ ^1\text{H} \ \rightarrow \ ^{13}\text{N} \ + \ \gamma$, and the ZAMS, marked as the moment when core hydrogen burning proceeds in full hydrostatic equilibrium. Several ZAMS definitions are found in the literature, for example, the ZAMS can be put after the settling of nuclear reactives, marked as a CN-burning bump, when nuclear burning accounts for at least 1\% of the total luminosity of the star \citep[e.g.,][]{Zwintz(2014)}. Another method sets the ZAMS according to the asteroseismic parameter $\Delta \nu$, that is, the distance in frequency between subsequent overtones of acoustic waves in stellar interiors, when this reaches 95\% of its maximum value on the MS \citep{Murphy(2023)}. The ZAMS can also be defined based on the central hydrogen abundance, marking the ZAMS as the moment it decreases by $\Delta X_\text{c}=0.01$ \citep[e.g.,][]{Steindl(2022):Impact_of_accretion}. In our work we opt for the latter abundance based definition, taking the ZAMS as the moment the central hydrogen abundance drops by 0.1\% from its initial value. Importantly, we choose to define the ZAMS using a relative percentage drop in $X_\text{c}$ rather than an absolute value, ensuring a consistent ZAMS definition across our mass range of 0.7 to 1.6 M$_\odot$. The ambiguity in ZAMS definitions is important when determining stellar ages, but does not change the overall conclusions of our work. As illustrated in left panel of Figure \ref{fig:HR_diagram_big}, our definition places the ZAMS some time after the depletion of $^{12}\text{C}$.

Pre-MS evolutionary tracks from 0.7 to 1.6 M$_{\odot}$ in increments of 0.2 M$_{\odot}$ are shown in the right panel of Figure \ref{fig:HR_diagram_big}. Ten models calculated with the classical prescription are drawn as gray lines, and the ten accreting models are drawn as colored lines. For this diagram, the accretion histories were hand-picked from the 35 histories we adopted from \cite{Elbakyan(2019)} to illustrate how different evolutionary paths can look depending on the accretion history.


The classical models are initialized as fully convective models of large radii that have already attained their full ZAMS mass. As they contract and heat up they descend down the Hayashi track until a radiative core develops, at which point they turn on to the Henyey track and later arrive on the ZAMS. The accreting models are instead initialized as second Larson cores which puts them at the bottom right of the HR-diagram. The extra heat injected into the models makes the earliest stages sensitive to the initial accretion rates and will affect the trajectories. Initial accretion rates can vary depending on the chosen accretion history, where the initial accretion rate corresponds to the first entry in the accretion history file. The trajectories also depend on the scaling applied to ensure that a model can reach the desired final mass as this can either decrease or increase the accretion rates (see Sec. \ref{subsec:mass_accretion_rates}). For example, the 0.7 M$_\odot$ model experiences an initial $\dot{M}$ of $10^{-5} \ \text{M}_\odot/\text{yr}$ under which it contracts and moves to hotter surface temperatures. The 1.3 M$_\odot$ model, on the other hand, experiences instead an initial $\dot{M}$ of $10^{-4} \ \text{M}_\odot/\text{yr}$ which results in a vertical trajectory towards higher luminosities. The subsequent evolution displays small fluctuations in both luminosity and temperature, reflecting the variable accretion rates. Occasional accretion bursts, with rates reaching up to $10^{-3} \ \text{M}_\odot/\text{yr}$, are also observed, manifesting as pronounced luminosity excursions in the HR-diagram. One such burst is observed in the 1.6 M$_\odot$ track (light blue line in right panel of Figure \ref{fig:HR_diagram_big}). Accretion continues until the ZAMS mass is reached, after which the models settle on tracks that resemble the classical ones, matching at the ZAMS. Our evolutionary tracks agree with results found in the literature \citep{Baraffe(2009), Kunitomo(2017), SJensen(2018), Elbakyan(2019), Steindl(2022):Impact_of_accretion}.

\subsection{Internal structure}
\begin{figure*}
    \centering
    \includegraphics[width=1.0\textwidth]{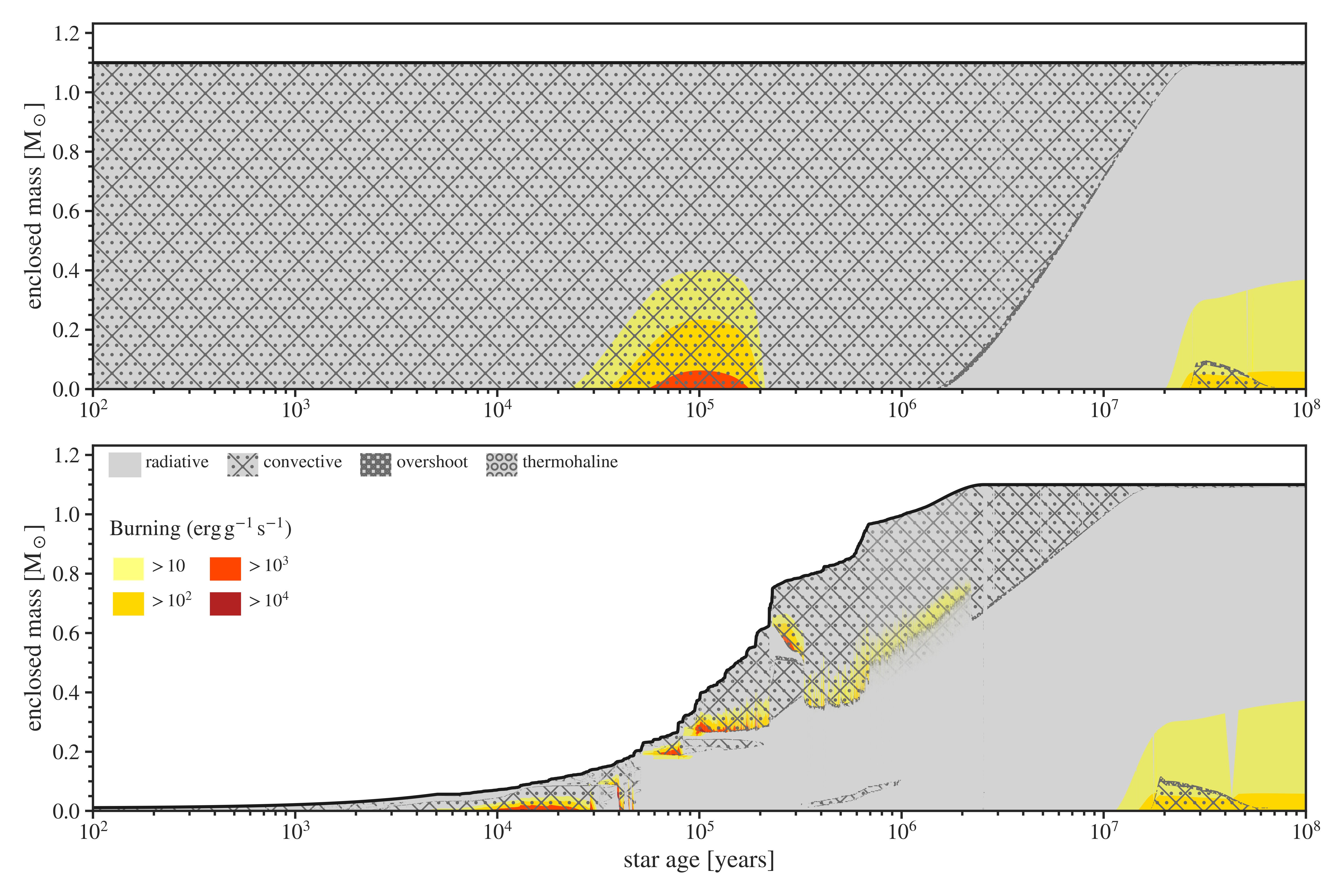}
    \caption{Kippenhahn diagrams for 1.1 M$_\odot$ classical (top) and accreting models (bottom). Radiative zones are colored gray and and zones of different mixing types are hashed. Burning strength is increasing from yellow to red.}
    \label{fig:kippenhahn}
\end{figure*}

The internal structures for 1.1 M$_\odot$ models in respectively the classical and the accreting scenarios are illustrated in Kippenhahn diagrams in Figure \ref{fig:kippenhahn}. Radiative zones are colored gray and zones of different mixing types are hashed. Zones are also colored from yellow to red according to burning strength. 

The two models are clearly distinguishable with the classical model maintaining constant mass and the accreting model gaining mass throughout its evolution. The classical model is fully convective, but gradually switches to radiative after $10^6$ years, and the depletion of deuterium occurs in the very center from contraction heating. In the accreting model the convective envelope and the radiative interior are firmly separated already around $5\cdot10^4$ years. Interestingly, this separation can have implications for surface abundances of heavy elements when the star arrives on the ZAMS. This is because composition changes in accretion disks may occur, for example, from the formation of planetesimals as such events are generally thought to deplete disks of metals \citep{Booth(2020),Huhn(2023)}.

Authors such as \cite{Serenelli(2011)} and \cite{Kunitomo(2021)} discussed these effects in the context of the solar abundance problem \cite{Asplund(2006)}; where modern solar models can't explain recently determined element abundances of metals at the solar surface. Specifically, utilizing pre-MS accretion, it is possible to build solar models with metal-rich interiors and a metal-poor envelopes, partially making up for the solar abundance problem. In our work, however, we do not concern ourselves with these effects as our models accrete material at a constant chemical composition representative of element abundances present in the original birth cloud. 

\begin{figure}
    \centering
    \includegraphics[width=1.0\linewidth]{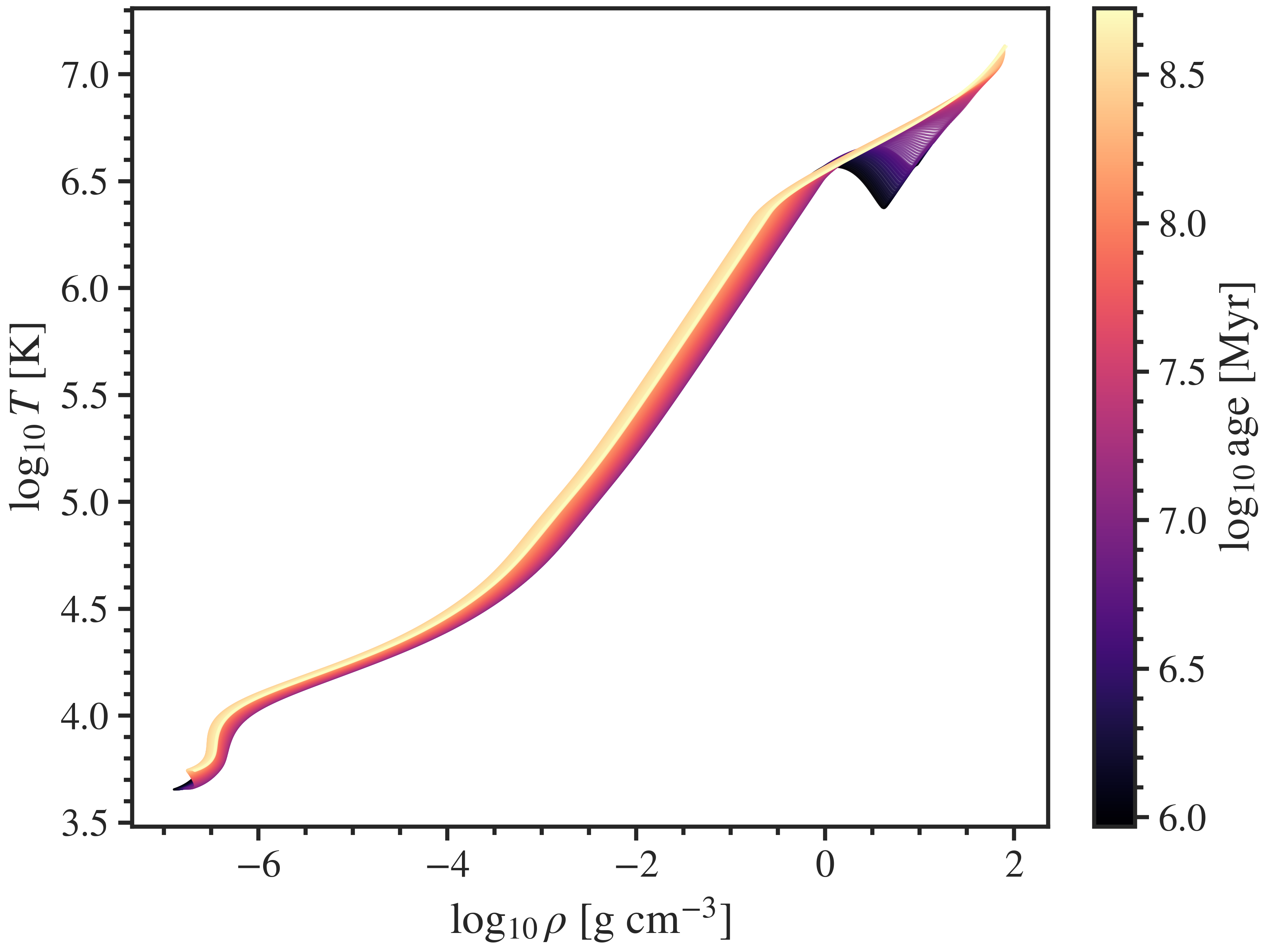}
    \caption{Temperature vs. density diagram for a one solar mass model. The profiles are colored by age and span a phase in which the model is at 90\% of its ZAMS mass up until the ZAMS. A temperature inversion is seen for the early stages of the model.}
    \label{fig:rho_vs_T}
\end{figure}

In addition to the separation, the accreting model also shows off-center burning at the base of the convective envelope throughout the accretion phase. In our scheme, strong mass accretion rates (bursts) lead to deep deposition of accretion energy and the presence of a temperature inversion from the core (illustrated in Figure \ref{fig:rho_vs_T}). Additionally, fresh deuterium from the accretion process is carried to the base of the convective envelope where it is ignited. This is the source of the off-center burning (red zones in the bottom panel of Figure \ref{fig:kippenhahn}). The initial deuterium depletion, however, still occurs at the center at around $1.5 \cdot 10^4$ years. 

The models resemble each other only after the accreting model has reached its final mass around $1.5 \cdot 10^6$ years, when no off-center burning is taking place and the subsequent evolution is characterized by a receding convective envelope. Similar trends are reported in the literature, for example, by \cite{SJensen(2018)} and \cite{Steindl(2022):Impact_of_accretion}.

\section{Asteroseismic imprint}

In this section we assess the imprint of the accretion processes on the asteroseismic properties of our models. In line with \cite{Steindl(2022):Impact_of_accretion}, we showcase altered buoyancy and Lamb profiles and quantify the magnitudes of frequency differences between classical and accreting models. Additionally, we replicate figures of an earlier study describing the theoretical existence of solar-like oscillations in pre-main sequence stars by \cite{Pinheiro(2008)}, namely their Christensen-Dalsgaard (C-D) diagram \citep{JCD(1993):CD-diagram} where the small frequency separation ($\delta\nu$) is plotted against the large frequency separation ($\Delta\nu$). Lastly, we perform a power-law fit between $\Delta\nu$ and $\nu_\text{max}$ values from our models.

\subsection{Propagation diagrams}
\begin{figure}
    \centering
    \includegraphics[width=1.0\linewidth]{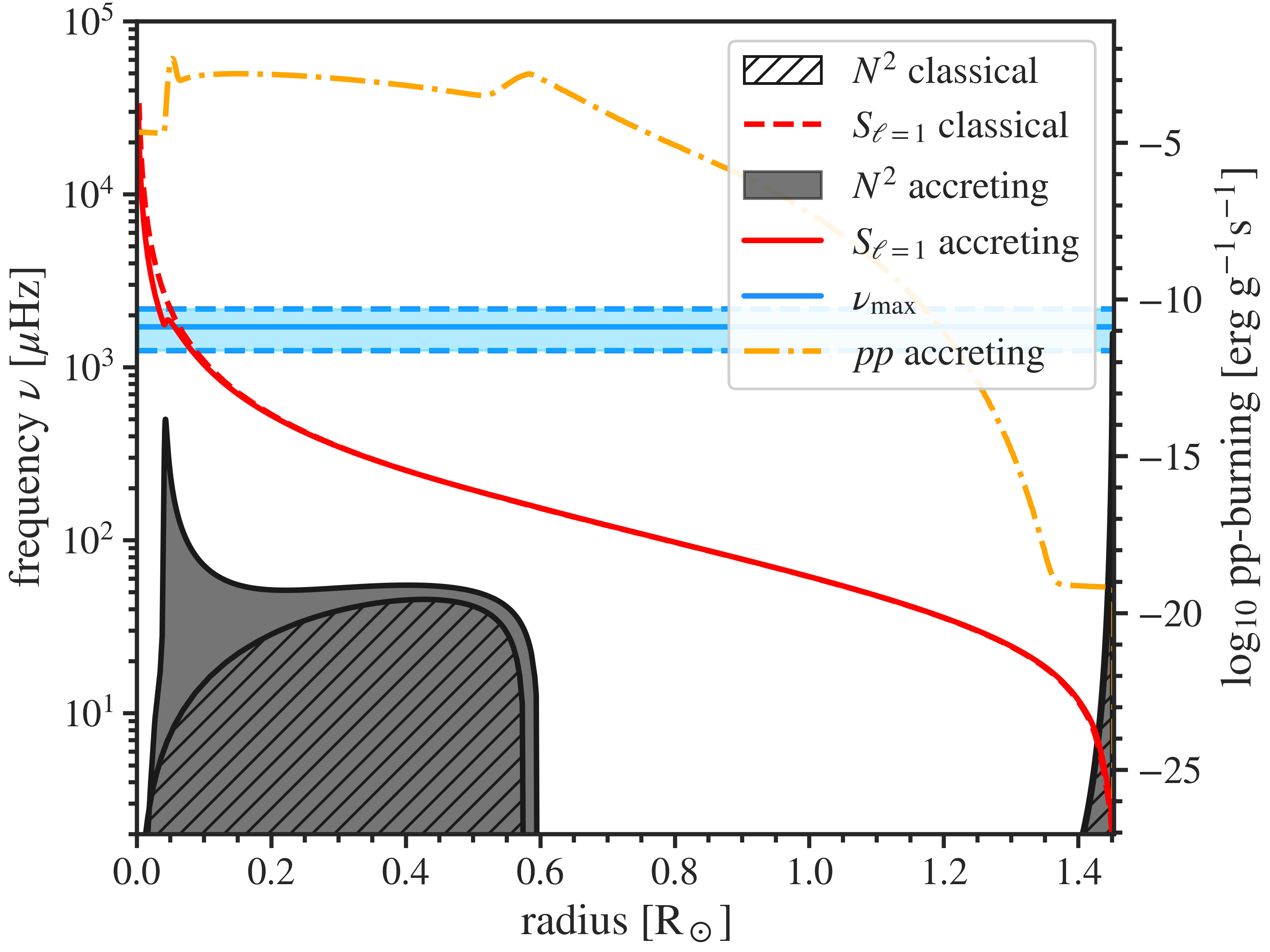}
    \caption{Propagation diagrams for 1.0 M$_\odot$ classical and accreting models. Buoyancy profiles are drawn in hashed for the classical model and in grey for the accreting model. The Lamb profiles of dipole modes are respectively drawn in dashed- and full red. The characteristic frequency, $\nu_{\text{max}}$, for the accreting model is also indicated along with the \textit{pp}-burning profile.}
    \label{fig:prop}
\end{figure}

To assess the asteroseismic imprint of accretion we draw in Figure \ref{fig:prop} propagation diagrams for a 1.0 M$_\odot$ classical model  an a 1.0 M$_\odot$ model that has recently emerged from the accretion phase. The models have similar radii and match each others positions in the HR-diagram (Figure \ref{fig:HR_classical_vs_accreting_prop}). They will therefore have similar $\nu_\text{max}$ values which is also indicated in the diagram (blue area). 

In propagation diagrams, internal oscillations cavities can be drawn against radius coordinates. A mode of oscillation will be of p-mode nature if it exists at frequencies above the buoyancy profile ($N^2$) and the Lamb profiles ($S_{\ell}$). There, it will propagate from the surface to an inner turning point marked by the Lamb profiles. The Lamb frequency profiles are given by
\begin{equation}
    S_\ell = \frac{\ell(\ell+1)}{r^2} c^2_\text{s},
\end{equation}
where $\ell$ is the spherical degree of a given mode when it is represented on the surface of a sphere with the spherical harmonics, $r$ is the turning point and $c_\text{s}$ is the sound speed. $\ell$ represents the number nodes on the surface in the longitudinal direction and for increasing $\ell$ we generally find modes of higher frequencies. Space-based photometry is limited by the effects of integrated light and modes with $\ell>3$ are not observed. Among the modes with $\ell\leq3$, the dipole modes ($\ell=1$) have the highest amplitudes (\citealp[see e.g.,][]{Lund(2017):LEGACY}). In Figure \ref{fig:prop} we therefore only draw the dipole Lamb frequency profile ($S_{\ell=1}$) for a given model. 

A mode existing solely within the buoyancy profile will be of g-mode nature, propagating from one edge of the profile to the other (\citealp[see e.g.,][]{Aerts(2021):Probing_the_interior_of_stars_with_asteroseismology}). Assuming a fully ionized ideal gas law for the equation of state, the buoyancy profile ($N^2$) can approximately be written as
\begin{equation}
    N^2 \simeq \frac{g^2\rho}{p} (\nabla_\text{ad} - \nabla + \nabla_\mu),
\end{equation}
with $g$, $\rho$ and $p$ being the local quantities of gravity, density and pressure. The expression in the parenthesis is essentially the Ledoux criteria for convective instability \citep{Ledoux(1947)}, with $\nabla_\text{ad}$ and $\nabla$ being the adiabatic and the actual temperature gradients respectively, and $\nabla_\mu$ the chemical gradient. These are formally written as \citep{Aerts(2021):Probing_the_interior_of_stars_with_asteroseismology},
\begin{equation}
    \nabla = \frac{\text{d} \ln{T}}{\text{d}\ln{p}},\quad \nabla_\text{ad}=\left(\frac{\partial \ln{T}}{\partial\ln{p}}\right)_S, \quad \nabla_\mu=\frac{\text{d} \ln{\mu}}{\text{d}\ln{p}},
\end{equation}
where $T$, $p$, and $\mu$ are local quantities of temperature, pressure and mean molecular weight, and $S$ denotes constant entropy. The extent of the buoyancy profile is therefore mainly determined by the depth of the convective envelope. This also means that when spurious convection zones occur the buoyancy profile can show gaps. In Figure \ref{fig:prop}, however, we selected models in an evolutionary stage where their buoyancy profiles are easier to compare.

We find differences in the buoyancy profiles between the two models (denoted $N^2$ classical and $N^2$ accreting in Figure \ref{fig:prop}). The profile of the accreting model has a sharp increase at small radii which we attribute to strong proton-proton (\textit{pp}) burning at this location which introduces a steep chemical gradient. At the same location the dipole Lamb profile of the accreting model (red solid line in Figure \ref{fig:prop}) shows a slight perturbation from the presence of the prominent buoyancy profile. Conversely, the cavities for the classical model are more separated and they will not influence each other. Evidence of this is seen in the smooth behavior of dipole Lamb profile for the classical model (red dashed line in Figure \ref{fig:prop}). The kink in the accreting model will affect the inner turning points of acoustic waves propagating in the stellar interior; consequently frequency differences between the two models are expected \citep{Steindl(2022):Impact_of_accretion}.  

\subsection{Evolution of frequencies}

\begin{figure}
    \centering
    \includegraphics[width=1.0\linewidth]{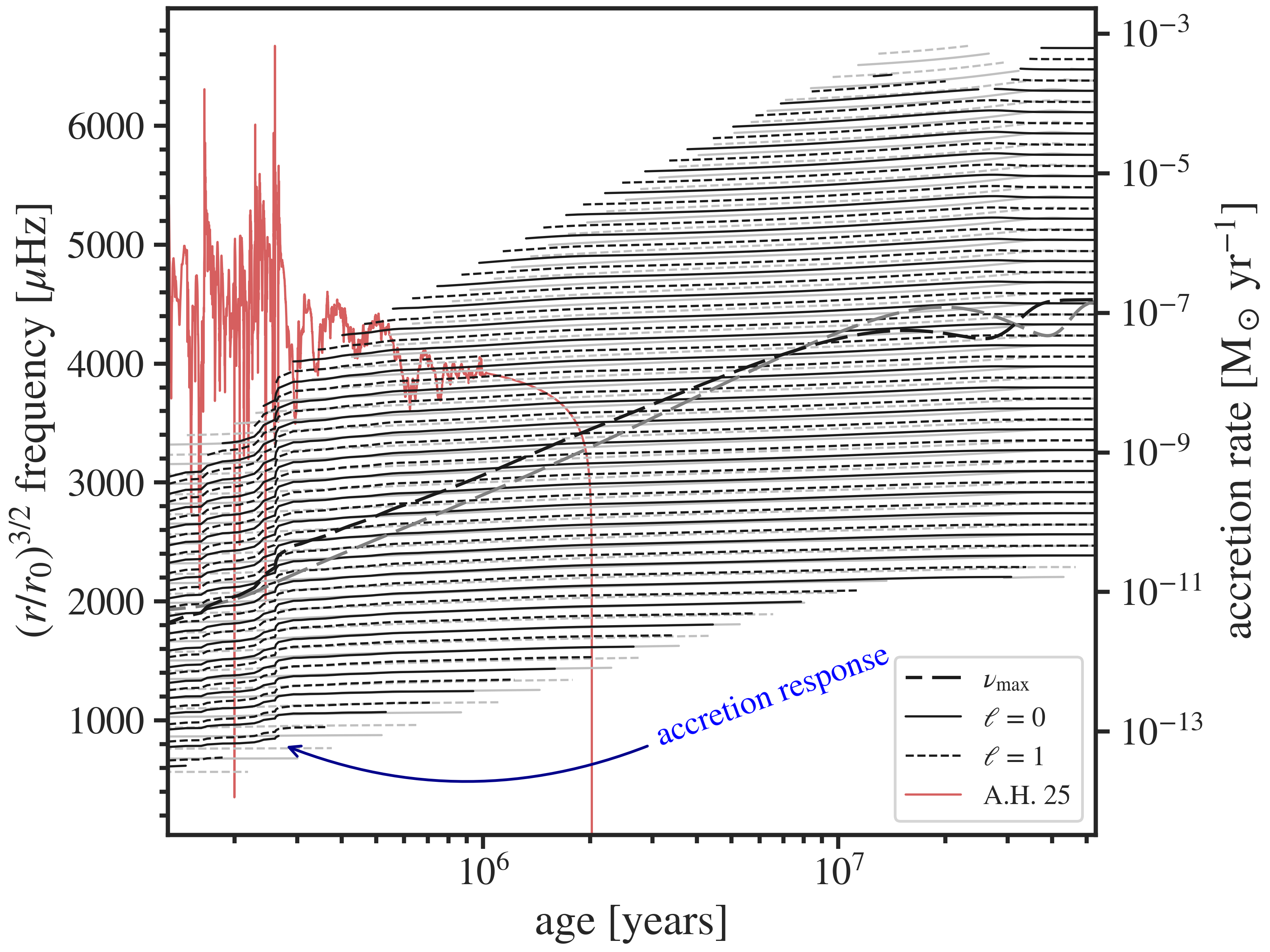}
    \caption{Evolution of frequencies for classical (gray) and accreting (black) 0.9 M$_\odot$ models. Radial modes are drawn as full lines, dipole modes as dashed lines, and $\nu_\text{max}$ as thicker dashed lines. The accretion history is drawn in red, and a blue arrow highlights the individual frequencies (black lines) responding to mass accretion. For visualization, the frequencies are scaled with third root of the ratio of the radius to ZAMS radius $(r/r_0)^{3/2}$.}
    \label{fig:compare_freqs_evolution}
\end{figure}

In the next step, we investigate the evolution of frequencies along the pre-MS. Figure \ref{fig:compare_freqs_evolution} illustrates the results exemplarily for a 0.9 M$_\odot$ classical model (gray) and a 0.9 M$_\odot$ accreting model (black). For better visibility the frequencies are scaled with the third root of the ratio of the radius to the ZAMS radius $(r/r_0)^{3/2}$ \citep{Hekker(2017)}. Radial modes are drawn as full lines, dipole modes as dashed lines, and $\nu_\text{max}$ as thicker dashed lines. The accretion history is also drawn in red.

Following the gray lines, illustrating frequencies in the classical approach, we observe $\nu_\text{max}$ to steadily increase, representing the continuous contraction along the Hayashi and Henyey tracks. A turning point in $\nu_\text{max}$ occurs as the model settles on the ZAMS (at an age of about $3.5 \cdot 10^7$ years). Following the black lines, illustrating the accreting model, similar trends in $\nu_\text{max}$ are evident. Differences emerge, however, in the individual frequencies. At ages smaller than one million years, when the model is still undergoing mass accretion, the frequencies are responding to the mass accretion. Once accretion dies down (after two million years), this behavior is not observed any more. Additionally, for the accreting model, the arrival on the ZAMS marks a similar turning point in $\nu_\text{max}$ as for the classical model. It also happens at about the same age as in the classical model, around $2\cdot10^{7}$ years. The results depicted in Figure \ref{fig:compare_freqs_evolution} describe the behavior of the frequencies from a theoretical perspective. Observations of solar-like oscillations during the accretion phases will be challenging as the signal of the accretion will likely dominate and obscure the subtle signal of the pulsations. On the other hand, accretion only makes up the first one million years of the entire pre-MS evolution. A star around one solar mass arrives on the ZAMS with an age already greater than tens of millions of years (see for example Figs. \ref{fig:kippenhahn} and \ref{fig:compare_freqs_evolution}). During this window the signal of accretion is mostly absent, with more periodic signals such as surface spots dominating the variability (see for example \cite{Cody(2014), Mullner(2021)}). Disentangling this signal from solar-like oscillations is feasible \citep{Corsaro(2024):Consecutive_smoothin}, provided that magnetic activity and rapid rotation do not fully suppress the oscillations.

\subsection{Frequency differences}
\begin{figure}
    \centering
    \includegraphics[width=1.0\linewidth]{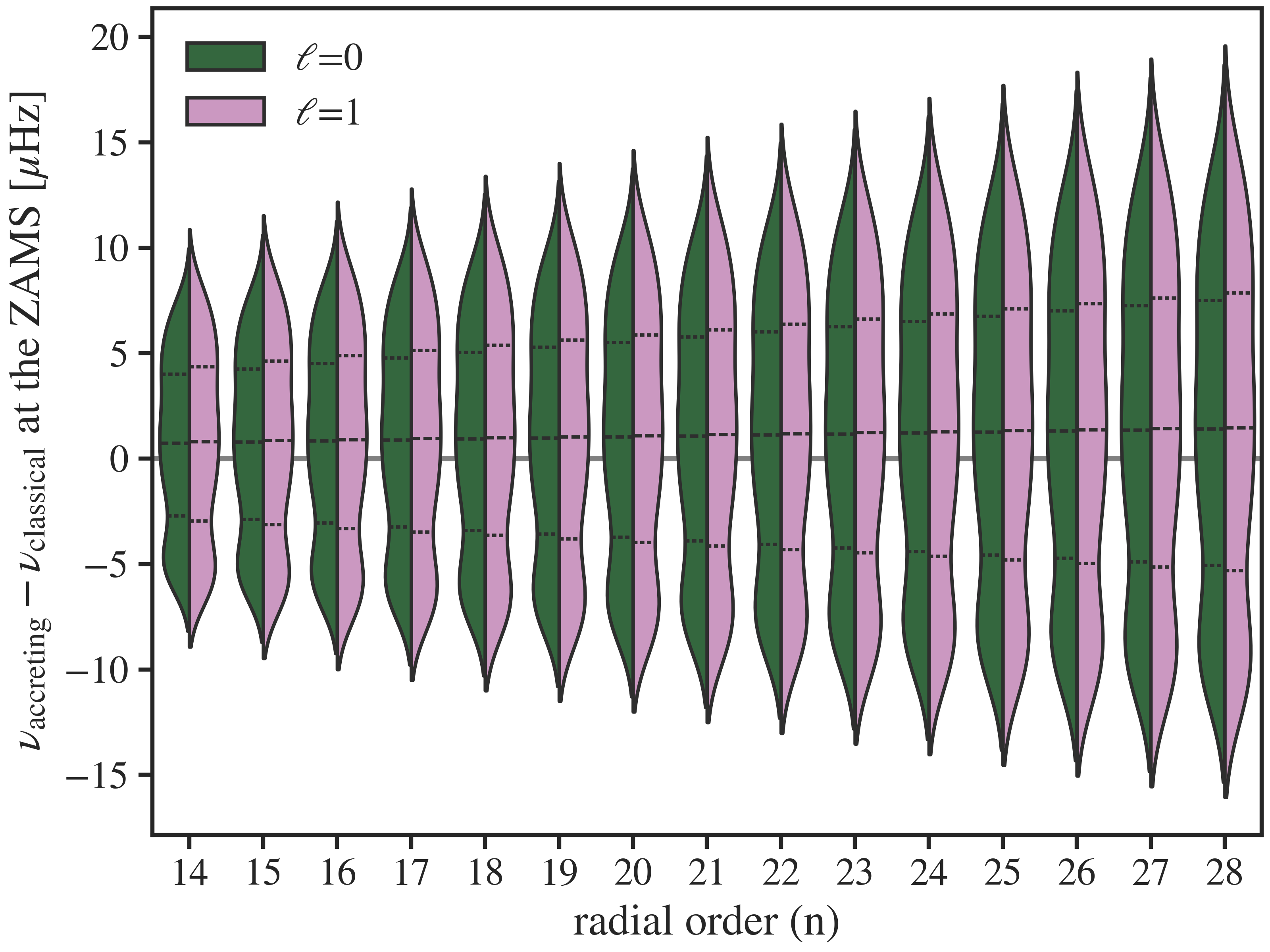}
    \caption{Violin diagram illustrating frequency differences at the ZAMS between classical and accreting models of one solar mass. Radial modes are shown in green and dipole modes in pink. Dashed lines indicate means of the distributions and dotted lines the 25th and 75th quartiles.}
    \label{fig:frequency_differences}
\end{figure}

As it might be challenging to detect solar-like oscillations in the accretion phase (up to a few million years) of these low-mass stars, we next investigate if the frequency differences depending on the description of the pre-MS evolution persist to a time period after the main accretion phase and until the ZAMS. In Figure \ref{fig:frequency_differences} we display the frequency differences at the ZAMS between classical and accreting models of 1.0 M$_\odot$ as violin plots.  Frequency differences are calculated per radial order ($n$), meaning that for a given mass (here 1.0 M$_\odot$) we calculate the differences between the 35 accreting models and a single classical model for modes of the same $n$. The violins thus illustrate how frequency differences between the 35 accretion histories are distributed per $n$. Differences in radial modes are drawn in green and differences in dipole modes in pink. Additionally, dashed lines indicate the mean value of the frequency differences, and the dotted lines indicate the 25th and 75th quartiles.

We find frequency differences between the accreting and the classical models, which increase with radial order. Positive frequency differences indicate that the classical models in general have smaller frequencies than the accreting models (see Figure \ref{fig:frequency_differences}) and vice versa. For the exemplary model of Figure \ref{fig:frequency_differences}, we find mean frequency differences around $0.7-1.5 \ \mu\text{Hz}$ with tails of the distributions reaching $-15$ and $+20$ $\mu\text{Hz}$ for the highest radial order ($n=28$). With approximately 12 quarters of Kepler data, that is, three years of continuous observations, uncertainties on individual frequencies can be on the order of $0.1-1.0 \ \mu\text{Hz}$ \citep[see e.g.,][]{Lund(2017):LEGACY}. Should future space missions reach similar precision for pre-MS stars, the frequency differences shown in Figure \ref{fig:frequency_differences} could play a role in fully unlocking the potential of pre-MS asteroseismology.

\subsection{C-D diagram}
\begin{figure}
    \centering
    \includegraphics[width=1.0\linewidth]{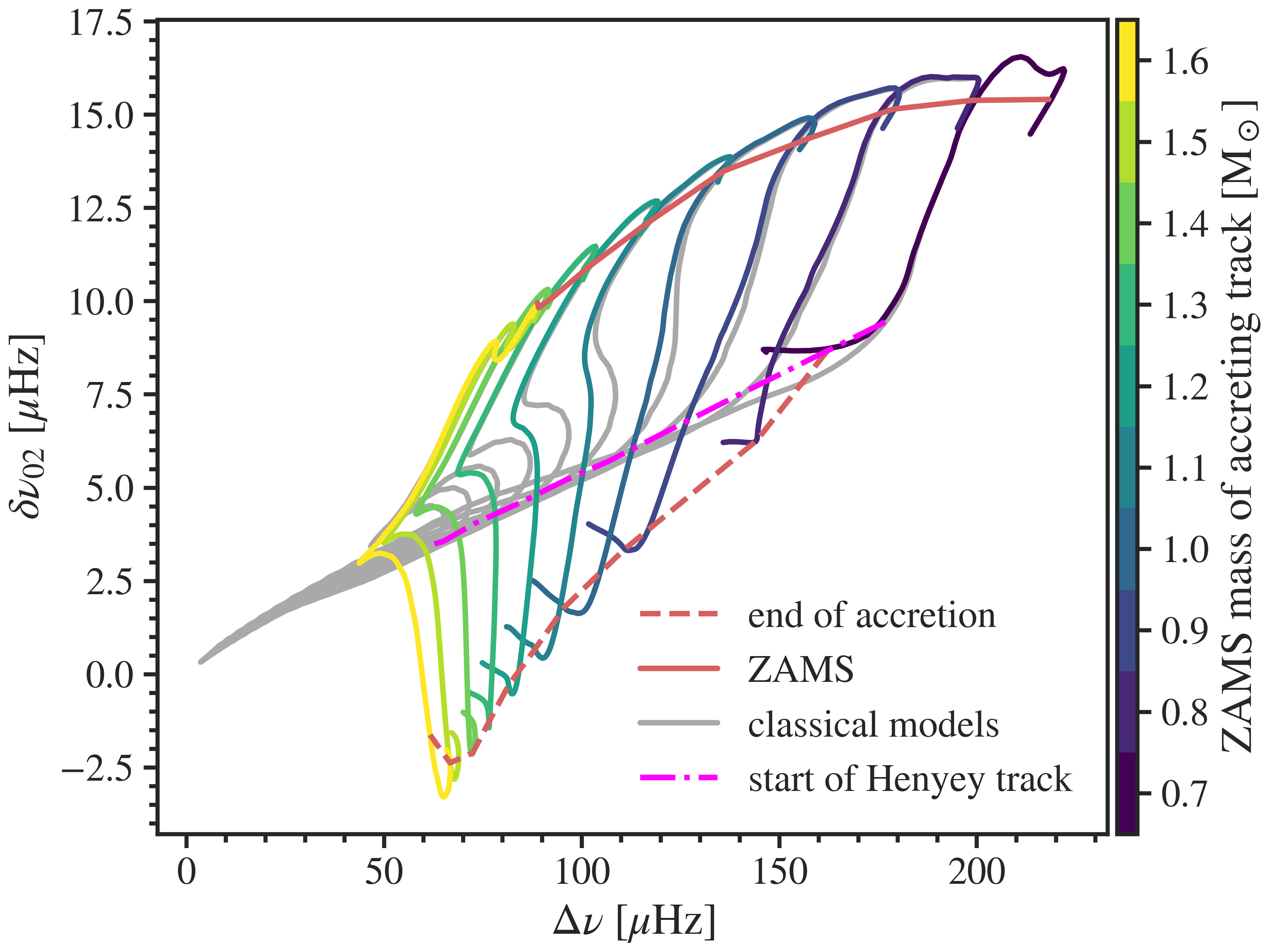}
    \caption{C-D diagram displaying the small separation against the large separation. Classical models are plotted in grey and accreting models are colored according to ZAMS mass, increasing from blue to yellow. The moment accretion ends is marked by a dashed red line, the ZAMS by a full red line, and the beginning of the Henyey track as a pink dot-dashed line.}
    \label{fig:CD_diagram}
\end{figure}

The Christensen-Dalsgaard diagram \citep[or C-D diagram;][]{Christensen_Dalsgaard(1988)} is a well-known asteroseismic diagram that relates the large separation, $\Delta\nu$, to the small separation, $\delta\nu$. The large frequency separation is calculated as the mean difference between all radial modes,
\begin{equation}\label{eq:mean_Dnu}
    \Delta\nu = \frac{1}{N} \sum_n^{n_\text{max}-1} (\nu_{n+1} - \nu_{n})_{\ell=0} \ ,
\end{equation}
and the small frequency separation as the mean difference between radial and quadrupole modes,
\begin{equation}\label{eq:mean_dnu}
    \delta\nu_{02} = \frac{1}{N} \sum_n^{n_\text{max}} \nu_{n, \ell=0} - \nu_{n, \ell=2} \ .
\end{equation}
In both equations $N$ is the total number of modes for a given $\ell$ that fall within the frequency envelope around $\nu_\text{max}$ for a given model, $n$ is the radial order and $n_\text{max}$ is the highest radial order for the modes calculated. $N$, $n_\text{max}$ and $\nu_\text{max}$ will evolve and vary with both mass and age, nevertheless, they could take the following values for an exemplary model considered in the analysis below (0.987 M$_\odot$, 1.625 Myr, accretion history nr. 3):
\begin{equation*}
    N=24, \quad n_\text{max}=35, \quad \nu_\text{max}=2304 \ \mu\text{Hz}.
\end{equation*}
The large separation reflects the overall properties of the star, like its mass, radius, and mean density. The small separation is more sensitive to the interior structure. C-D diagrams commonly also connect tracks by lines of constant hydrogen abundance which evidently could not be realized for our pre-MS models as no full equilibrium hydrogen burning is taking place. In Figure \ref{fig:CD_diagram} we show the C-D diagram for classical models (grey lines) and accreting models (colored according to ZAMS mass, increasing from blue to yellow) calculated with accretion history number three. We highlight the end of accretion with a dashed red line and the ZAMS with a full red line. Finally, to distinguish Hayashi and Henyey tracks of the classical models, we draw the beginning of the Henyey track (when there is a turning point in luminosity) as a pink dot-dashed line.

It can clearly be seen in Figure \ref{fig:CD_diagram} that pre-MS tracks in the C-D diagram are well separated to the point where individual masses can be distinguished. The classical models are all collapsed for small values of $\Delta\nu$ and $\delta\nu_{02}$ which represents the contraction along the Hayashi track. Our results are consistent with earlier work by \cite{Pinheiro(2008)} who noted that C-D diagrams offer little diagnostics for pre-MS stars above 1.3 M$_\odot$. When considering models already on the Henyey track (beginning of which is marked as a pink dot-dashed line), we reach similar conclusions as we observe that curves converge as we go to higher masses. However, during the phase where our models have attained the ZAMS mass but not yet reached the Henyey track, curves are more distinguishable even as we go beyond 1.3 M$_\odot$. This behaviour challenges the findings of \cite{Pinheiro(2008)}, as we find that pre-MS solar-like stars can be distinguished in the C-D diagram during the immediate post-accretion phase.

\subsection{$\Delta\nu$-$\nu_\text{max}$ scaling relation}

\begin{figure}[h!]
    \centering
    \includegraphics[width=1.0\linewidth]{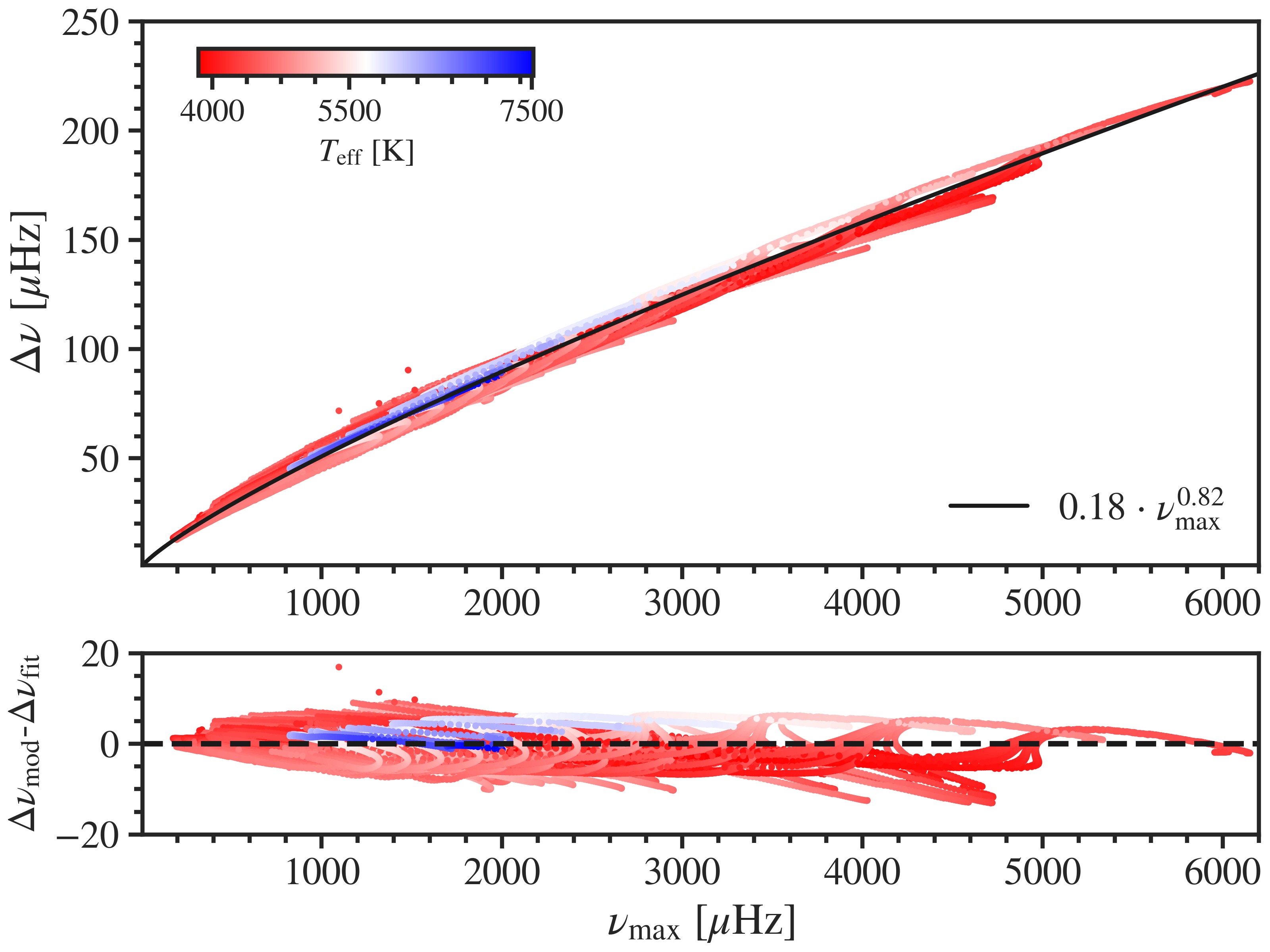}
    \caption{Top panel illustrates a power-law fit between $\Delta\nu$ and $\nu_\text{max}$ values derived from our models (top) and the corresponding residuals (bottom). A 1:1 line is drawn in black, and the data points are additionally colored by temperature increasing from red to blue colors. Bottom panel illustrates residuals with the zero-line drawn in dashed black.}
    \label{fig:dnu_numax_scaling_relation}
\end{figure}

Powerful relations between the large frequency separation, $\Delta\nu$, and the frequency of maximum power, $\nu_\text{max}$, have been established for solar-like oscillators in evolutionary stages beyond the pre-MS \citep{Stello(2009), Huber(2011):Scaling_relations}. These relations provide a way to estimate $\Delta\nu$ from a known $\nu_\text{max}$ that is derived from the analysis of observational data. $\Delta\nu$ is calculated as in equation (\ref{eq:mean_Dnu}) and our best guess for $\nu_\text{max}$ in pre-MS stars follows the scaling relations as explained in Sec. \ref{subsec:GYRE}. In Figure \ref{fig:dnu_numax_scaling_relation} we show the relation between $\Delta\nu$ and $\nu_\text{max}$ based on our accreting class II (Sec. \ref{subsec:GYRE}) pre-MS models.

We fit a power-law with an exponent of 0.82 which aligns well with literature values that are typically between 0.76 and 0.82 \citep{Stello(2009), Huber(2011):Scaling_relations}. Fits are usually performed separately for stars on the MS and the red giant branch (RGB) \citep{Huber(2011):Scaling_relations}. It is possible that a similar distinction might be necessary for different phases during the pre-MS, for example, distinguishing the Henyey track from the approach to the ZAMS where the models show an inversion in $\Delta\nu$. As this is subject of future work, however, we do no make such a division here and observe residuals smaller than 20 $\mu$Hz. An additional feature of Figure \ref{fig:dnu_numax_scaling_relation} is that we find $\nu_\text{max}$ values mainly greater than 500 $\mu$Hz. Given the Nyquist limits of 30 minutes long-cadence spaced-based photometry, it can be argued that such data are not suitable to resolve the expected frequencies of young solar-like oscillators.

\begin{figure}[h!]
    \centering
    \includegraphics[width=1.0\linewidth]{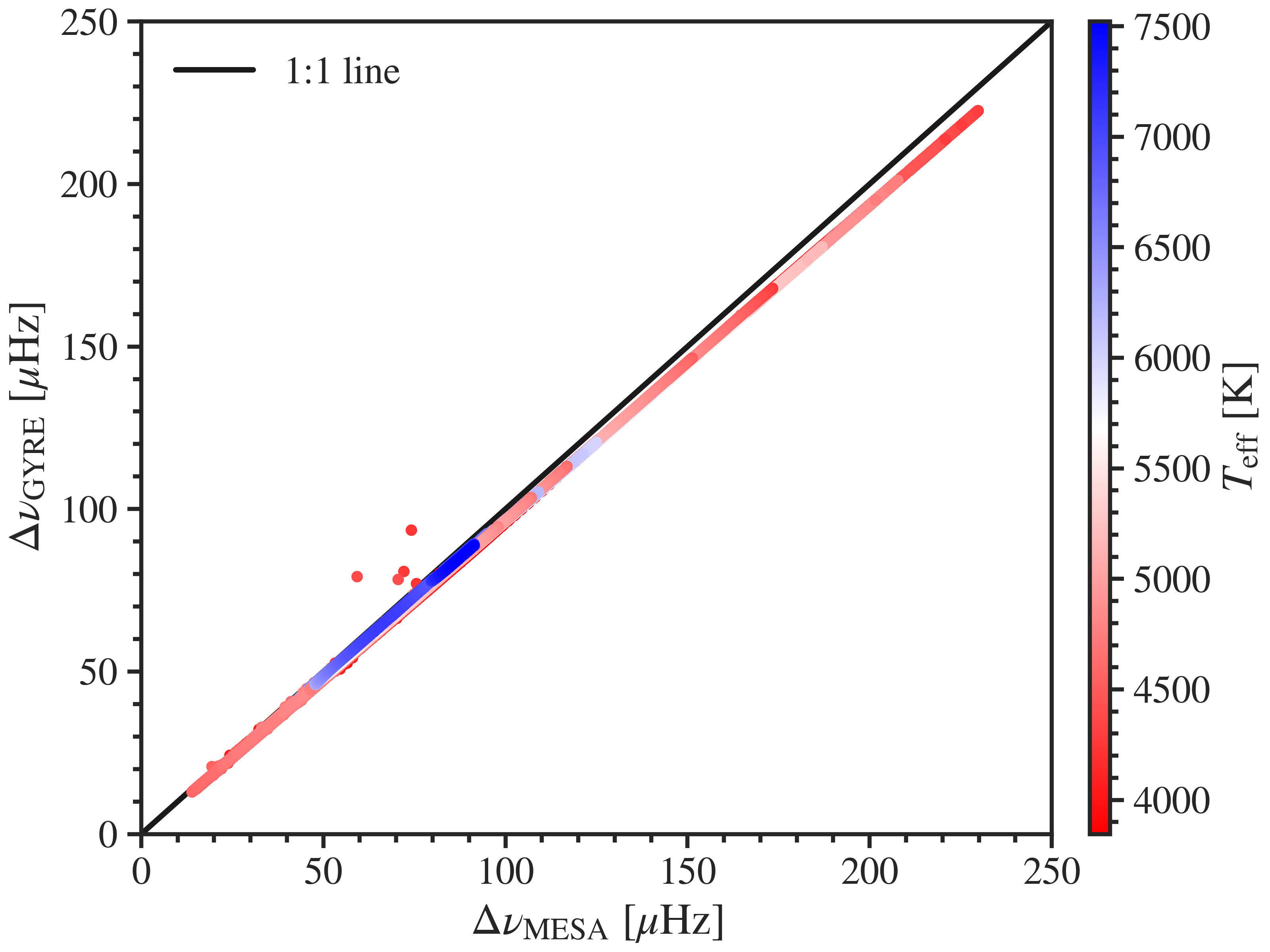}
    \caption{$\Delta\nu$ values obtained from \texttt{GYRE} ($\Delta\nu_{\texttt{GYRE}}$) drawn against $\Delta\nu$ values derived from \texttt{MESA} ($\Delta\nu_{\texttt{MESA}}$). Each data point is colored by $T_\text{eff}$ increasing from red to blue colors, and an ideal one-to-one line is also drawn in black.}
    \label{fig:mesa_vs_gyre_dnu}
\end{figure}

The validity of the scaling relations can be assessed by comparing $\Delta\nu$ values obtained from \texttt{GYRE}, computed using equation (\ref{eq:mean_Dnu}), with those derived from \texttt{MESA}, calculated as
\begin{equation}\label{eq:Dnu_MESA}
    \Delta\nu = \left(2\int_{0}^{R} \frac{\text{d}r}{c_\text{s}(r)}\right)^{-1},
\end{equation}
which is the inverse of twice the sound travel time between the center and the surface of a stellar model \citep[see, e.g.,][]{Aerts(2021):Probing_the_interior_of_stars_with_asteroseismology}. We argue that if the $\nu_\text{max}$ values derived from the atmospheric parameters of our models, using equation (\ref{eq:numax}), are reliable, then the $\Delta\nu$ values from \texttt{GYRE} and \texttt{MESA} should be mutually consistent. This is illustrated in Figure \ref{fig:mesa_vs_gyre_dnu} where $\Delta\nu_{\texttt{GYRE}}$ is drawn against $\Delta\nu_{\texttt{MESA}}$. Each data point is colored by $T_\text{eff}$ increasing from red to blue colors. For context, an ideal one-to-one line is also drawn in black.

We find that $\Delta\nu$ values are mostly consistent for $\Delta\nu<100 \ \mu\text{Hz}$, but increasingly deviate beyond $100 \ \mu\text{Hz}$. At this point, we can offer no real explanation for this behavior, but we note that the agreement is overall acceptable. A real calibration of the scaling relations for pre-MS stars requires more confirmed detections \citep{Mullner(2021)}, hence the the already established relations serve as our best guess for $\nu_\text{max}$ values of pre-MS stars.

\section{Preparing for PLATO}
PLAnetary Transits and Oscillation  of stars \citep[PLATO; ][]{Rauer(2025)} is a medium-class European Space Agency (ESA) mission with an expected launch in December 2026. The primary science goals of PLATO (i.e., its Core Science Program) focus around the discovery and characterization of Earth-like planets orbiting solar-like stars in their habitable zones. PLATO has a multi-telescope design and carries in total 26 cameras -- 24 in white light and one camera each utilizing a red and a blue filter -- that cover a sky area of 2132 deg$^2$ \citep{Rauer(2025)}. Over its nominal mission duration of four years, the PLATO space telescope will monitor about 250 000 stars brighter than $V = 15$ mag. It is expected that PLATO will deliver photometric time series data of pristine quality not only for its Core Science targets, but for a multitude of different types of variable stars. For all scientific topics that can be addressed with PLATO data, but are different from the Core Science, the PLATO Complementary Science Program was designed \citep{Tkachenko(2024),Aerts(2024)}.

The first field of PLATO observations, LOPS2, was selected to be in the southern hemisphere \citep{Nascimbeni:LOPS2} overlapping the TESS continuous viewing zone and intersecting the Galactic plane -- the birthplace of stars in the Milky Way. Consequently, PLATO will also be able to observe pre-MS stars in LOPS2.

The pre-MS solar like stars we discuss here belong to the Complementary Science of PLATO. But at the same time they are the progenitors of PLATO's Core Science targets on the stellar side. Hence they are of particular interest to study the complete evolution of solar-like stars.
With the launch of the PLATO mission upcoming in a few months, the description of the pulsational characteristics of pre-MS solar like oscillators based on current models of early stellar evolution is quite relevant. Below, we therefore first explore the theoretical pulsational properties of pre-MS solar like stars and then discuss the observational challenges to detect confirmed members of this class in general.

\subsection{Data from previous and ongoing missions}
Currently, no confirmed detection of solar-like oscillations in pre-MS exist. 
Although previous missions have observed young stars, the time bases of their light curves are limited to a maximum of $\sim$100 days. 

The original Kepler mission \citep{Boruck(2010):Kepler} deliberately avoided star-forming regions, resulting in no time-series for pre-MS stars. The Kepler K2 mission \citep{Gilliland(2010):K2} did target young regions \citep{Rebull(2018)}, but observations were usually conducted with 30-minute cadence and delivered time bases of approximately 80 days. \cite{Mullner(2021)} analyzed K2 data and identified one candidate pre-MS solar-like oscillator. The NASA mission TESS \citep{Ricker(2014):TESS} conducts a nearly all-sky survey and has also targeted young regions allowing for several studies of the variability of pre-MS stars, for example, by \citet{Elizabethson(2023)} and \citet{Buoma(2024)}. Most TESS observations of young stars are in 30-minute cadence; only a limited number of targets are observed with a shorter cadence. The longest time bases available for photometric time series of pre-MS stars amount to $\sim$100 days and were recorded by the TESS mission. Due to the relatively large pixel scale of $21'' \times 21''$ of the TESS cameras, observations in the galactic plane and in star forming regions sometimes suffer from strong contamination issues.

\subsection{Detectability of pre-MS solar-like oscillations with PLATO data}
Based on our accreting models we expect the p-mode frequencies of solar-like oscillations in pre-MS stars to be in the range from $\sim 500$ to $6500\,\mu$Hz (see e.g., Figure \ref{fig:dnu_numax_scaling_relation}). Similarly to MS solar-like stars, the amplitudes of the pre-MS counterparts can be expected to be at the parts-per-million level. However, stellar activity components such as rotation and magnetic fields can further dampen the pulsations \citep{Chaplin(2000):Magnetic_dampening, Garcia(2011), Jenkins(2011), Bonanno(2014), Zwintz&Steindl(2022), Corsaro(2024), Bessila(2024)}. PLATO photometric time series will have time bases of at least two years, which already more than doubles the currently available lengths for pre-MS stars, enabling the detection of smaller amplitude signals. Observations will be carried out in short cadences of 25 and 6.25 seconds \citep{Rauer(2025)} allowing to resolve the theoretically predicted solar-like frequencies. Additionally, the PLATO cameras have a pixel scale of $15'' \times 15''$ which is smaller than the TESS pixel scale. Crowding issues should be less dramatic compared to TESS.

\subsection{Pre-MS stars in LOPS2}
\begin{figure}
    \centering
    \includegraphics[width=1.0\linewidth]{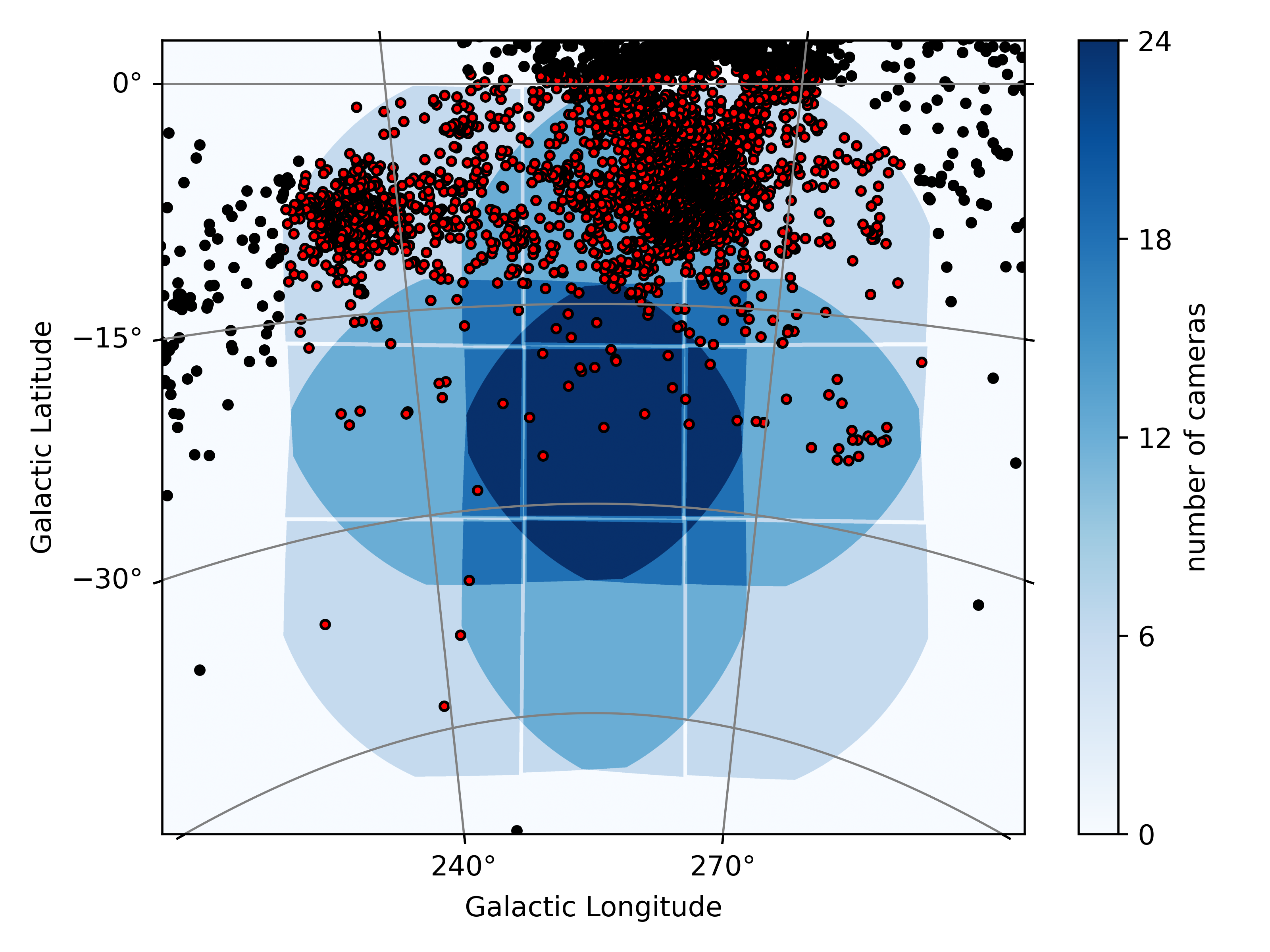}
    \caption{Young stars from SPYGLASS IV \citep{Kerr(2023)} for which $P_{\text{Age}<50 \  \text{Myr}} > 0.99$ and that fall within LOPS2 (red symbols). The stars are concentrated to the upper part of the field-of-view which is where PLATO intersects the galactic plane. Stars marked as black symbols do not fall within LOPS2} 
    \label{fig:nplatocam}
\end{figure}

The first field that PLATO will observe, LOPS2, is centered at the equatorial coordinates RA$_{\rm 2000} = 16^h \, 28^m \, 43.2^s$ and DE$_{\rm 2000} = -47^{\circ} \, 53' \, 13''$. It contains star forming regions, young open clusters and known young stellar objects (see Figure \ref{fig:nplatocam}). 

\begin{figure*}
    \centering
    \includegraphics[width=1.0\textwidth]{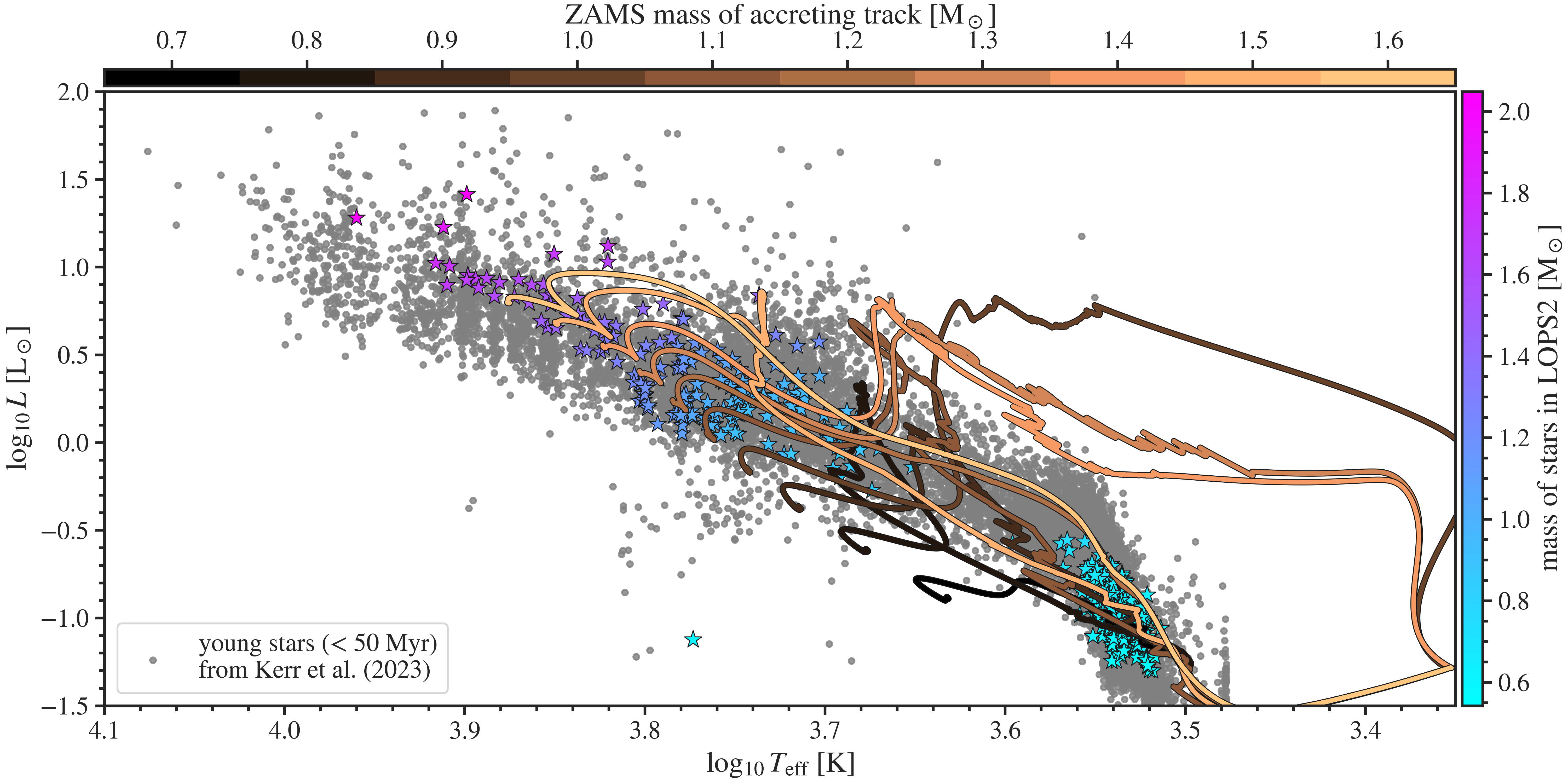}
    \caption{Young stars from the SPYGLASS survey by \protect\cite{Kerr(2023)} with ages less than 50 million years. Grey symbols represent the whole sample while colored symbols represent stars in LOPS2 for which masses are available. Luminosities, effective temperatures and masses are taken from \protect\cite{Creevey(2023)}. Evolutionary tracks are also plotted, increasing in mass from black to orange colors.}
    \label{fig:spyglass}
\end{figure*}

Figure \ref{fig:spyglass} shows stars deemed photometrically young from the SPYGLASS IV survey by \cite{Kerr(2023)} and for which the ages are estimated to be less than 50 million years at 99\% confidence (grey symbols, $P_{\text{Age}<50 \  \text{Myr}} > 0.99$). The authors deem $P_{\text{Age}<50 \  \text{Myr}} > 0.2$ a sufficient youth indication, however, given the size of the survey, we decided to be more strict in our criteria. The study of \cite{Kerr(2023)} additionally concerns stars within 1 kpc of the Sun. Of the 32835 targets in our sample, 3552 are located within LOPS2. For 688 of these stars, luminosities, effective temperatures and masses are available from the Gaia astrophysical parameters inference system \citep{Creevey(2023)}. For context, selected theoretical, accreting evolutionary tracks are also included.

We identify young low-mass stars in LOPS2 with masses in the range where their main-sequence counterparts would show solar-like oscillations. The observed luminosities, effective temperatures, and masses are compatible with our stellar models. In the bottom right of Figure \ref{fig:spyglass} the data points do not intersect the ZAMS which is expected as such stars do not reach the ZAMS until roughly 100 million years after formation.

\subsection{Complications in detections}
\label{sec:complications}
Confirming solar-like oscillators during the pre-MS stages is complicated by several factors. Fast rotation rates and magnetic fields dampen the oscillations \citep{Chaplin(2000):Magnetic_dampening, Garcia(2011), Jenkins(2011), Bonanno(2014), Corsaro(2024), Bessila(2024)}. Surface spots introduce signal modulation at low frequencies. The latter are usually removed by means of filtering the light curves (see for example \citealp{Mullner(2021), Corsaro(2024)}). Additionally, pre-MS stars are faint with Gaia and TESS magnitudes typically greater than 14 \citep{Elizabethson(2023), Kerr(2023)}. Detections of solar-like oscillations in faint stars, for which the noise level is large, show promising results with Gaussian processes though \citep{Hey(2024)}. Moreover, pre-MS stars often reside in crowded regions where custom apertures perform better than pipelines (see for example \citealp{Steindl(2022)II}).

\begin{figure}
    \centering
    \includegraphics[width=1.0\linewidth]{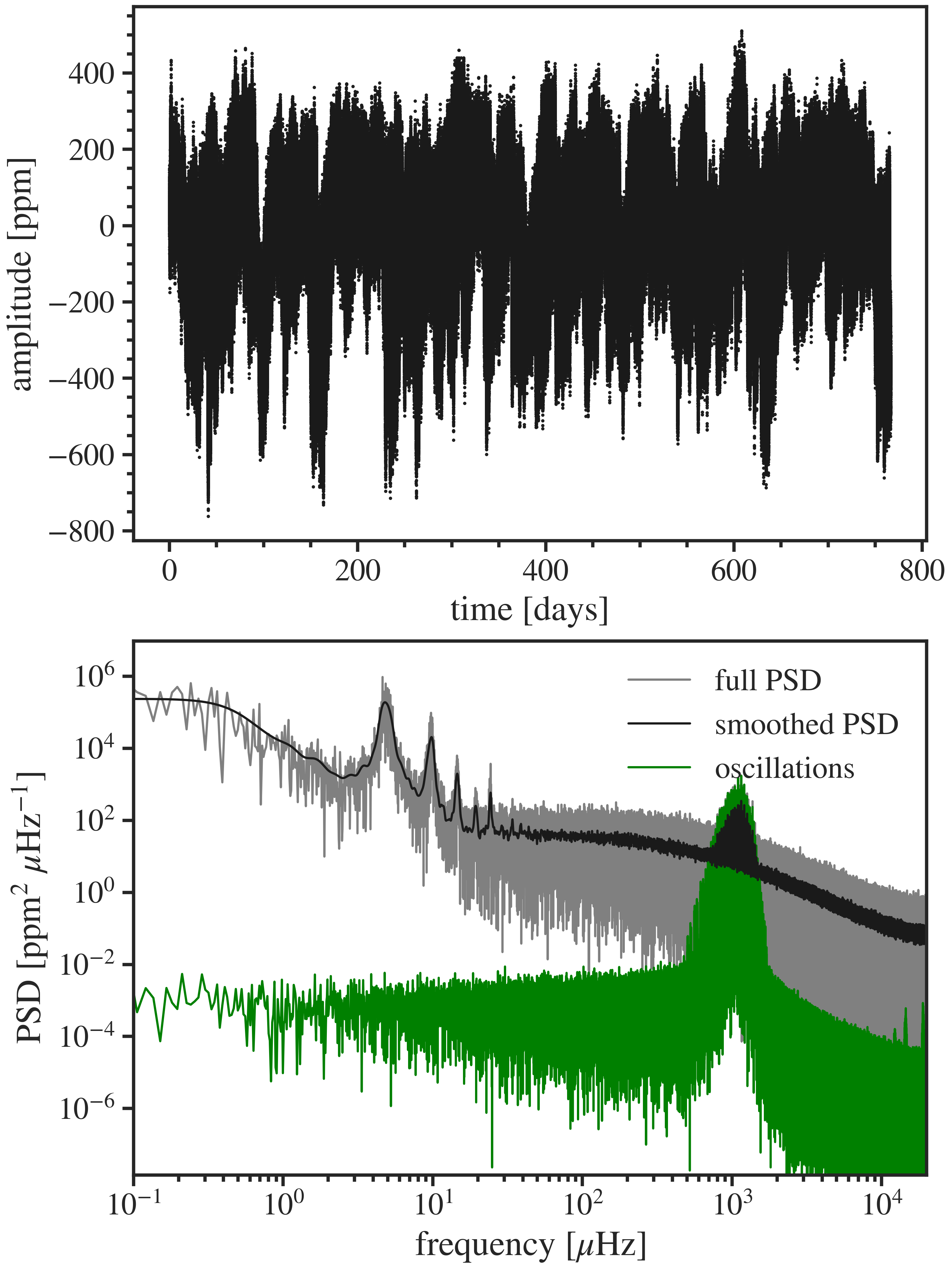}
    \caption{Simulated light curve (top) for a young pre-MS star exhibiting spot modulation and solar-like oscillations. The corresponding power spectrum (bottom) is drawn in grey with a filtered spectrum in black. The pure oscillation signal is drawn in green}
    \label{fig:simulation}
\end{figure}

Using our models and Gaia data we simulate light curves of pre-MS solar-like stars to further discuss the potential for their detections from future observations. A simulated noise-free light curve and power spectrum are shown in Figure \ref{fig:simulation}. The simulated light curve is two years long in accordance with the monitoring of LOPS2 and shows variability from surface spots. This manifests as modulation at low frequencies in the power spectrum where the pure oscillation signal is shown in green. The alias peaks above 10 000 $\mu$Hz result from the Fourier transform and are not physical. The simulations of the oscillation signal and granulation background follow the formalism of \cite{DeRidder(2006)} with frequencies taken from our \GYRE models and realistic granulation and oscillation parameters from \cite{Kallinger(2014)} and \cite{Lund(2017):LEGACY}. Spot modulation is added using the software \texttt{Butterpy} \citep{Butterpy(2022)} with rotation rates taken from \cite{Elizabethson(2023)}. A dampening of the power excess is added according to an approximate formula from the work of \cite{Bessila(2024)}; as an inverse power-law, $A_{\text{damp}}=(1+P_{\text{rot}}/P_{\text{rot},\odot})^{-0.25}$, where $A_\text{damp}$ is the dampening factor, $P_{\text{rot}}$ is the rotation period and $P_{\text{rot},\odot}=28$ is the rotation period of the Sun in days. Figure \ref{fig:simulation} illustrates an ideal case of 2-years long PLATO observations as the simulations are noise-free, but it also demonstrates that the comparatively faint pulsation signal can be detected in the presence of a strong signal originating from spots on the stellar surface. Future work will include a fully realistic model of the PLATO cameras into our simulations through PLATOsim \citep{Jannsen(2024)}. Our light curve simulator is available on GitHub\footnote{\url{https://github.com/Johanneshj/pre-ms-solar-like-oscillators-light-curve-simulator}}.

\section{Conclusions}

In this work we have presented low-mass pre-MS stellar models undergoing initial mass accretion until their ZAMS mass is reached. Accretion physics significantly increases the dimensionality of stellar models. For that reason, we focused only on the effects of hot and variable accretion as this is required to produce excursions in the HR-diagram from accretion bursts as manifested by FU-Orionis eruptions, which we deemed of interest. Effects we have not accounted for include rotation (young stars are known to be rapid rotators, see, for example, \citealp{Elizabethson(2023)}), metal rich and poor accretion \citep{Serenelli(2011)}, or even a metallicity dependent accretion rate \citep{Kunitomo(2021)}. The accreting evolution is inherently different from the classical scenario where models are initialized at their ZAMS mass with the subsequent evolution segmented into Hayashi and Henyey tracks. We compared evolutionary tracks and internal structures between the two descriptions. Asteroseismic properties of the models were examined with \GYRE. Finally, we discussed our findings in the context of the PLATO mission. Our main conclusions are as follows:

Accretion processes significantly alter evolutionary tracks of pre-MS stars. This is in line with previous studies who demonstrated, for example, that the luminosity spread in open clusters can be explained by episodic accretion at early stages of evolution \citep{Baraffe(2009),SJensen(2018)}. Through the scaling relation, the frequencies of solar-like oscillators are sensitive to atmospheric parameters such as luminosity, effective temperature and surface gravity (equation \ref{eq:numax}). The parameter space we find for accreting low-mass stars should then better represent what we can expect for $\nu_\text{max}$ of these objects.

The internal structures of pre-MS stars are altered by accretion processes. We have illustrated in Kippenhahn diagrams that the deposition of accretion energy in the envelope leads to off-center nuclear burning. This is in contrast to the classical scenario where contraction heating is the source of burning during the pre-MS. Mixing regions are also affected, with the accreting models developing radiative zones much earlier than the classical. The models are compared also in propagation diagrams where differences in the buoyancy profiles are observed. The asteroseismic imprint of mass accretion in stellar models is discussed in a similar manner in \cite{Steindl(2022):Impact_of_accretion} and \cite{Wagg(2024)}.

Mass accretion affects the asteroseismic properties of pre-MS models. We have shown that individual frequencies respond rapidly to variable accretion rates and that the imprint of accretion carries on through to the ZAMS where frequency differences are observed. We quantified the obtained frequency differences in violin diagrams, illustrating frequency differences across the 35 formation scenarios adopted for this study. For the model comparisons in Figure \ref{fig:frequency_differences}, we find frequency differences smaller than $20 \ \mu\text{Hz}$ at the ZAMS. We have also shown that C-D diagrams are different when effects of accretion are included. Specifically, in the near-ZAMS phase individual tracks can be distinguished up until $1.3 \ \text{M}_\odot$. This degeneracy for higher masses is not present in the immediate post-accretion phase where individual tracks are distinguishable across all masses we considered. Finally, we performed a power-law fit between $\Delta\nu$ and $\nu_\text{max}$ values from our models, finding an exponent of 0.82 \citep{Stello(2009), Huber(2011):Scaling_relations}, and also examined the validity of the scaling relations for pre-MS stars. 

PLATO will provide data for young solar-like stars. We illustrated that young solar-like stars fall within the field of view of PLATOs first field LOPS2. The mission will thus bring about two years continuous monitoring of these objects which will increase the chances of catching the small signal of solar-like oscillations. We have begun work to quantify the known complications with observing young stars (rotation, magnetic fields, accretion disks, crowding) with simulated data. This will serve as a first estimate on the asteroseismic yield of young solar-like stars in PLATO.

It is worthwhile to discuss the implications of the differences in asteroseismic properties between classical and accreting models. In some sense, including formation histories can be thought of as adding yet another dimension to the already large dimensionality of stellar models (for example chemical composition, $\alpha_\text{mlt}$, initial masses). With the work presented here, however, we have shown that when accretion is not included it is possible that calculated frequencies carry uncertainties that are not properly propagated in stellar modeling techniques. This was tackled in the modeling of the pre-MS $\delta$-Scuti star HD 139614 in \cite{Steindl(2022)II}, however, as shown in \cite{Murphy2023} and \cite{ZhaoGou(2024)}, frequency differences stemming from different formation scenarios are usually smaller than differences observed from other effects such as rotation or even choices of nuclear reaction networks. Asteroseismic modeling of p-mode pulsators mainly make use of extracted frequencies, and the question of whether it makes sense to invest computational effort in calculating accreting pre-MS models becomes valid. Even so, as we and previous studies have shown, it is now evident that the pre-MS is more complex than previously thought. \citep{Siess(1997), Baraffe(2009), Vorobyov(2015), Kunitomo(2017), SJensen(2018), Elbakyan(2019), Steindl(2021)I, Steindl(2022)II, Steindl(2022):Impact_of_accretion, Zwintz&Steindl(2022)}. To the extent that accretion processes can be implemented in one dimensional stellar evolution codes, widely different evolutionary tracks and internal structures are observed in pre-MS models. The effects of, for example, off-center deuterium ignition are not captured in classical models. With ever increasing data quality and continuous refinement of stellar modeling techniques, the subtle effects of different formation scenarios could play a key role in unlocking the full potential of asteroseismology of pre-MS stars.

\begin{acknowledgments}
We thank the developers of \MESA and \GYRE for providing and maintaining the publicly available stellar structure and stellar pulsation codes. We also thank members of the \MESA community, in particular Ebraheem Farag, Jared Goldberg, and Pablo Marchant for their help with \MESA specifics on the \MESA user mailing list. We thank Andrés Ramirez for insightful comments on our analysis. E.I.V. acknowledges
support from the FWF project I4311-N27. The stellar models were computed using the LEO5 high-performance computing cluster at the University of Innsbruck and we gratefully acknowledge the computational resources provided by the University of Innsbruck IT Services. This work has made use of data from the European Space Agency (ESA) mission Gaia (\url{https://www.cosmos.esa.int/gaia}), processed by the Gaia Data Processing and Analysis Consortium (DPAC, \url{https://www.cosmos.esa.int/web/gaia/dpac/consortium}). Funding for the DPAC has been provided by national institutions, in particular the institutions participating in the Gaia Multilateral Agreement. This work has also made use of Matplotlib, a graphics package for Python for publication-quality image generation \citep{Hunter(2007)}; NumPy \citep{VanDerWalt(2011)}; MESA SDK for Linux (Version 22.6.1, \citealp{Townsend(2020):SDK}); Astropy \citep{astropy:2013, astropy:2018, astropy:2022}; SciPy \citep{2020SciPy-NMeth}.
\end{acknowledgments}

\begin{contribution}

Johannes Jørgensen was responsible for calculating the stellar models and producing the figures, as well as writing and submitting the manuscript. The research concept was outlined by Konstanze Zwintz who also edited the manuscript. Thomas Steindl, Ebraheem Farag, and Eduard Vorobyov provided feedback on the original manuscript. Thomas Steindl aditionally provided the initial \MESA inlists, and the mass accretion histories were based on work by Eduard Vorobyov.


\end{contribution}

%



\appendix

\section{Numerical convergence}
In this section we discuss numerical convergence in our models and perform a resolution test. Future work will address the issues outlined below.

\subsection{Obtaining numerical convergence}
Here we explain choices made to obtain numerical convergence for our \MESA models. We identify the most problematic cases as either high accretion rates on to the initial stellar seeds, or during strong accretion bursts in the later evolution. In both cases the accretion timescale exceeds or is on the order of the thermal timescale ($\tau_{\dot{M}}\geq\tau_{\text{therm}}$). In the literature, such issues have been circumvented by tweaking \MESA's $\texttt{eps\_mdot\_leak\_frac}$ parameter \citep{Posydon} or by adopting the \MESA option to adopt the actual temperature gradient ($\nabla_T$) in cases where the use of hydrostatic $\nabla_T$ from mixing-length theory is inappropriate\footnote{\url{https://docs.mesastar.org/en/22.11.1/reference/controls.html\#use-gradt-actual-vs-gradt-mlt-for-t-gradient-eqn}}. The latter option is used in \cite{WongBildsten(2023)} who also noted better convergence for models experiencing mass loss when setting \MESA's $\texttt{eps\_mdot\_factor}$ to zero which neglects the redistribution of energy due mass changes \citep{Paxton2019}. 

\begin{figure}
    \centering
    \includegraphics[width=1.0\linewidth]{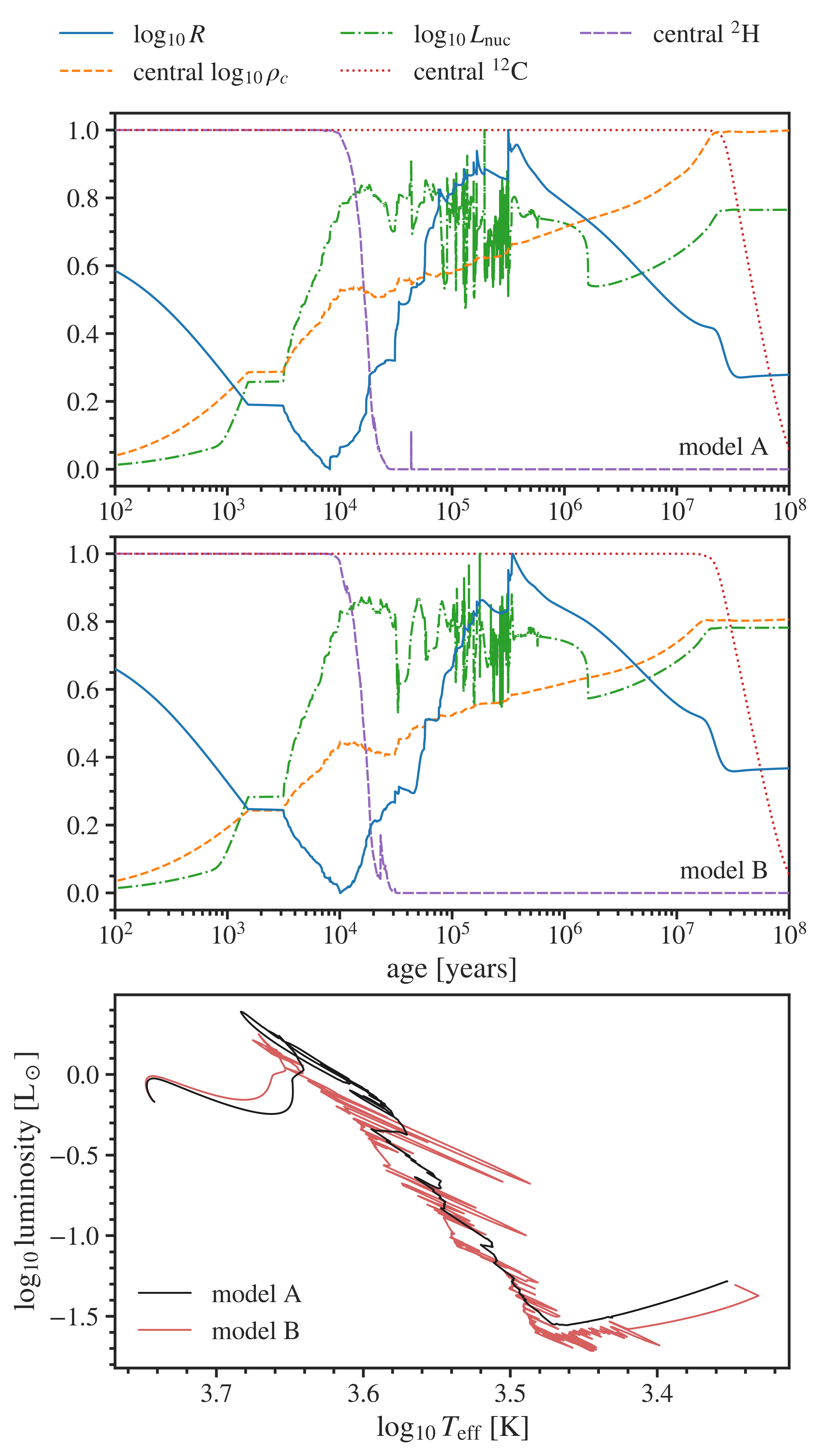}
    \caption{Comparison of 1.0 M$_\odot$, accretion history 35, models calculated with the two different \MESA setups, A (top) and B (middle), as described in the text. The diagrams display fundamental stellar parameters along the protostellar and the pre-MS phases. The bottom panel compares evolutionary tracks.}
    \label{fig:app:compare}
\end{figure}

In our work we adopt the dynamical settings for $\nabla_T$ in \MESA by enabling 
\begin{align*}
&\texttt{use\_gradT\_actual\_vs\_gradT\_MLT\_for...} \\
&\texttt{...\_T\_gradient\_eqn} = \texttt{True},
\end{align*}
which relates $\nabla_T$ to the actual temperature gradient rather than the one produced from the mixing length theory. We furthermore set $\texttt{eps\_mdot\_factor}=0$, which should be done with caution. Our choices are, however, motivated by the need to handle a wide range of accretion scenarios and stellar masses while minimizing the need for recalculations. With these settings enabled only a handful of models require restarts, in which case we relax the initial hard limits imposed on parameters such as luminosity and density, and then rerun the models. By setting $\texttt{eps\_mdot\_factor}=0$ we essentially neglect energy changes associated with mass changes, and our models are effectively pushed toward the cold accretion scenario in which the accretion process does not significantly heat the star and most of the released energy is radiated away \citep{Baraffe(2009),Baraffe(2012),SJensen(2018), Elbakyan(2019)}. The accretion routine outlined in section \ref{subsec:modelling_accretion} partly makes up for this simplification by manually injecting heat into the models. As a sanity check we compare two \MESA models where for one we adopt the settings above (model A) and for the other the standard \MESA settings are used (model B). The results of this test are shown in Figure \ref{fig:app:compare} which illustrates the differences between the two model descriptions across various parameters. The top two panels show the surface radius ($R$), central density ($\rho_c$), luminosity from nuclear fusion processes ($L_\text{nuc}$), central carbon-12 abundance (central $^{12}\text{C}$) and central deuterium abundance (central $^{2}\text{H}$) against age. All quantities have been normalized to the range [0, 1]. The inset text indicates which model descriptions are used and the bottom panel compares the evolutionary tracks of the two models. 

We see differences in surface parameters such as luminosity and effective temperature as well as in $L_{\text{nuc}}$. The remaining structural quantities exhibit broadly consistent behavior across both model setups, albeit with slight differences. Notably, both models consistently exhibit off-center deuterium burning during the accretion phase, a feature of particular interest. With this in mind, we should emphasize that it would not be appropriate to claim that our models with complete accuracy represent accreting low-mass pre-MS stars. On the other hand, numerical issues aside, the exact nature of star formation and deposition of accretion energy (Sec. \ref{subsec:modelling_accretion}) is not yet fully understood, and our models would have looked different had we opted for another scheme. Addressing the numerical problems in our \MESA setup will be an important focus of future work. Even so, while the \MESA settings used here were chosen to ensure grid coverage, our main conclusions remain robust under this setup; the main one being that accretion affects the asteroseismic properties of pre-MS stars. Our study thus provides a broad understanding of pre-MS solar-like oscillators and enables the calculation of representative oscillation spectra for the synthetic light curves discussed in Sec. \ref{sec:complications}.

\begin{figure}
    \centering
    \includegraphics[width=1.0\linewidth]{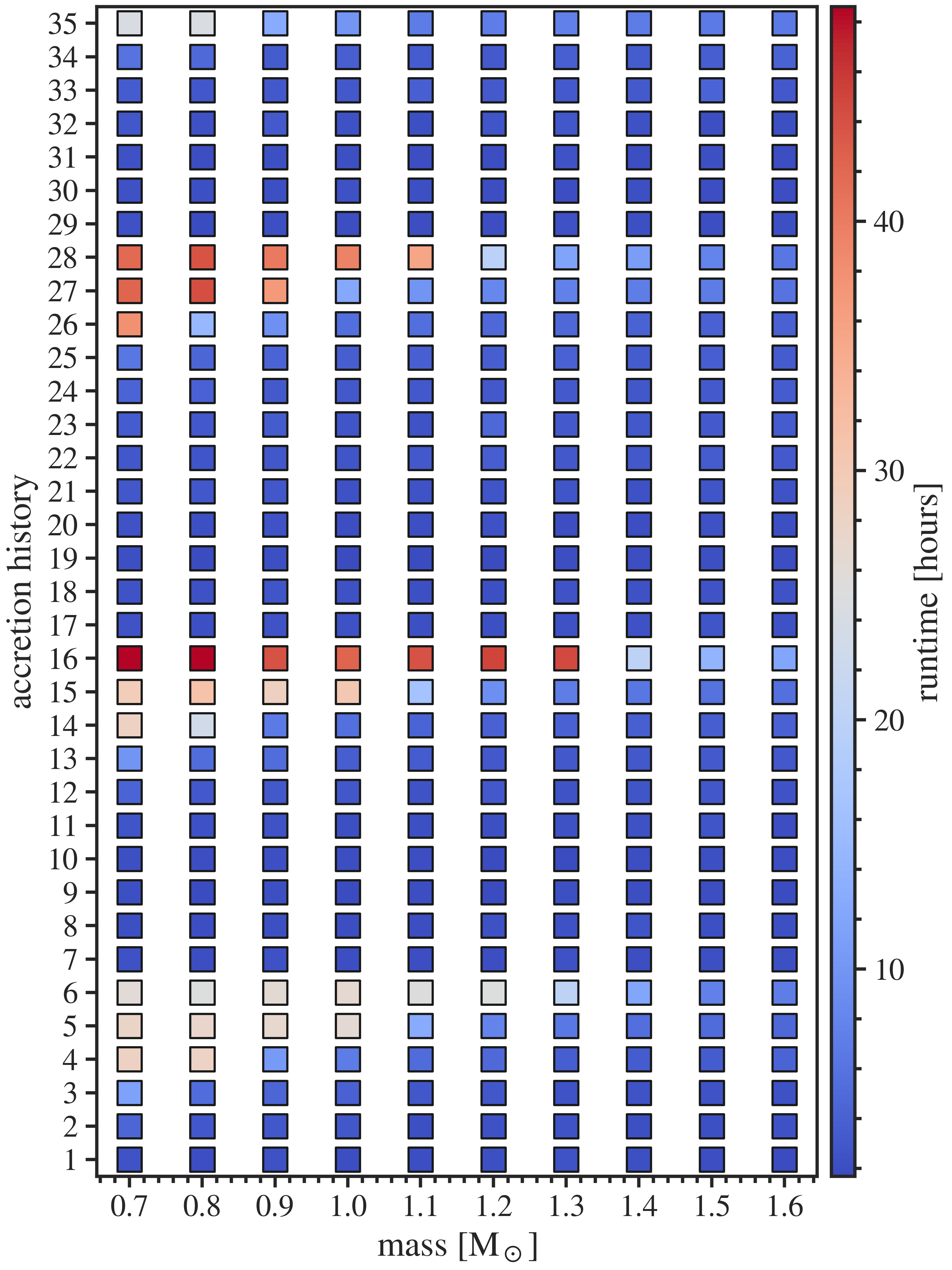}
    \caption{Convergence map for the calculated model grid. Squares indicate converged models and are colored by run time in hours increasing from blue to red.}
    \label{fig:app:convergence}
\end{figure}

Figure \ref{fig:app:convergence} demonstrates convergence and run times on six CPU cores across our model grid adopting the solutions explained above. Squares represent converged models and are colored by calculation time increasing from blue to red colors. 

\subsection{Resolution test}

\begin{figure}
    \centering
    \includegraphics[width=1.0\linewidth]{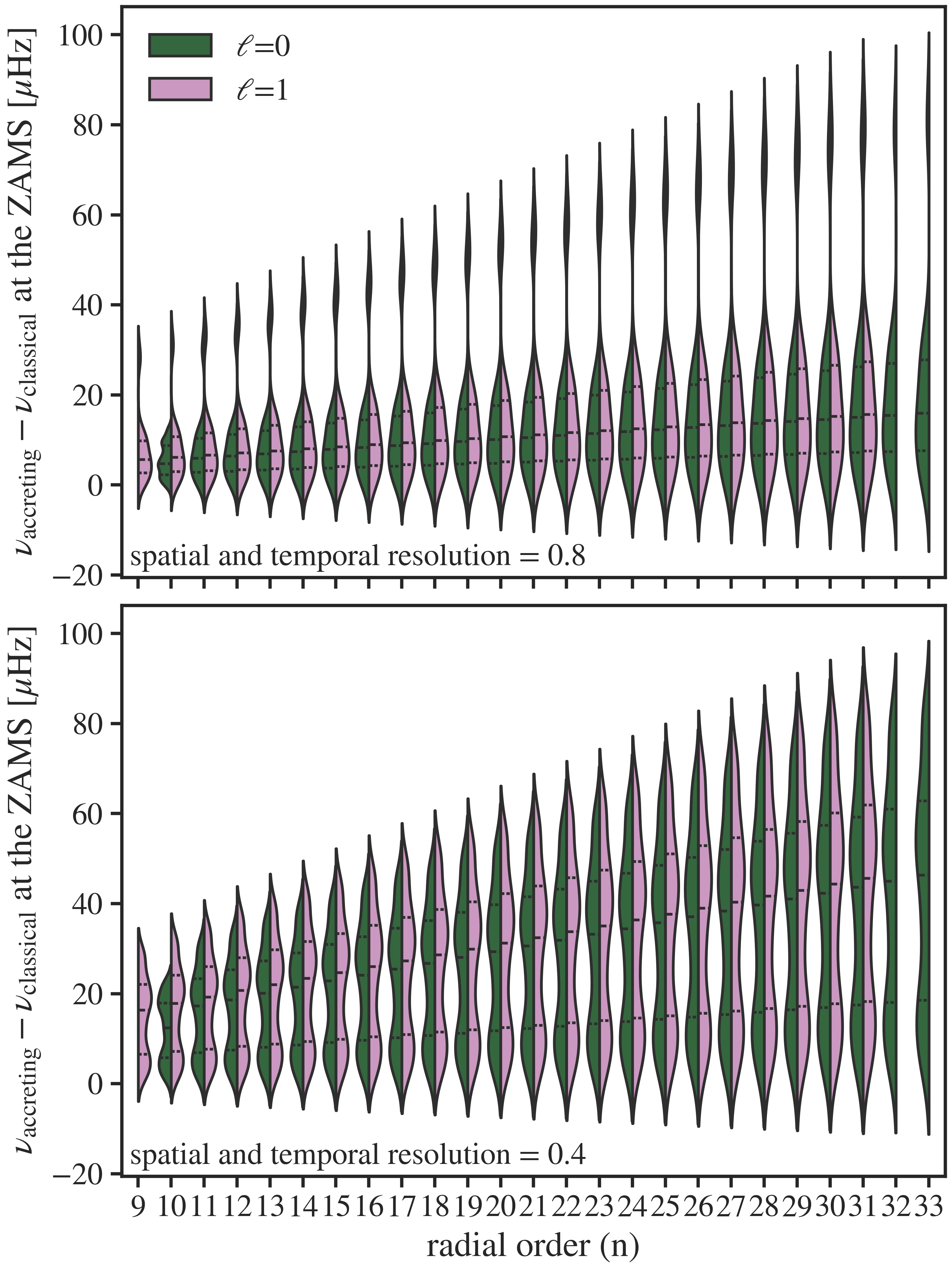}
    \caption{Frequency differences between resolutions of 0.8 and 0.4 for 1.2 M$_\odot$ models. Radial modes are shown in green, and dipole modes are shown in pink. The top panel corresponds to the coarser resolution, while the bottom panel corresponds to the finer resolution.}
    \label{fig:app:resolution_test}
\end{figure}

To test the extent to which our conclusions would change with increased resolution we perform a resolutions test. The figures in the main text are produced using models with the following \MESA settings for temporal and spatial resolution: $\texttt{time\_delta\_coeff}=0.8$ and $\texttt{mesh\_delta\_coeff}=0.8$. Here, we calculate a smaller set of models with these values reduced to 0.4, effectively doubling the resolution. This set of models is calculated for 1.2 M$_\odot$ and all 35 accretion histories. Figure \ref{fig:app:resolution_test} illustrates frequency differences between the two resolutions. The top and bottom diagrams correspond to resolutions of 0.8 and 0.4 respectively. Frequency differences for radial modes are drawn in green, while differences for dipole modes are shown in pink.

We find bimodality in the distributions for both resolutions, with this feature being more prominent at the finer resolution. This effectively pushes the average frequency differences for each radial order towards higher values. Even so, the envelopes traced by the minima and maxima of the distributions remain consistent across both resolutions. This indicates that, while small differences exists between the two resolutions, increasing them will not affect our overall conclusions.

\section{Supplementary figures}\label{app:sec:supplementary_figures}

Here we present supplementary figures to support the main text.

\begin{figure}
    \centering
    \includegraphics[width=1.0\linewidth]{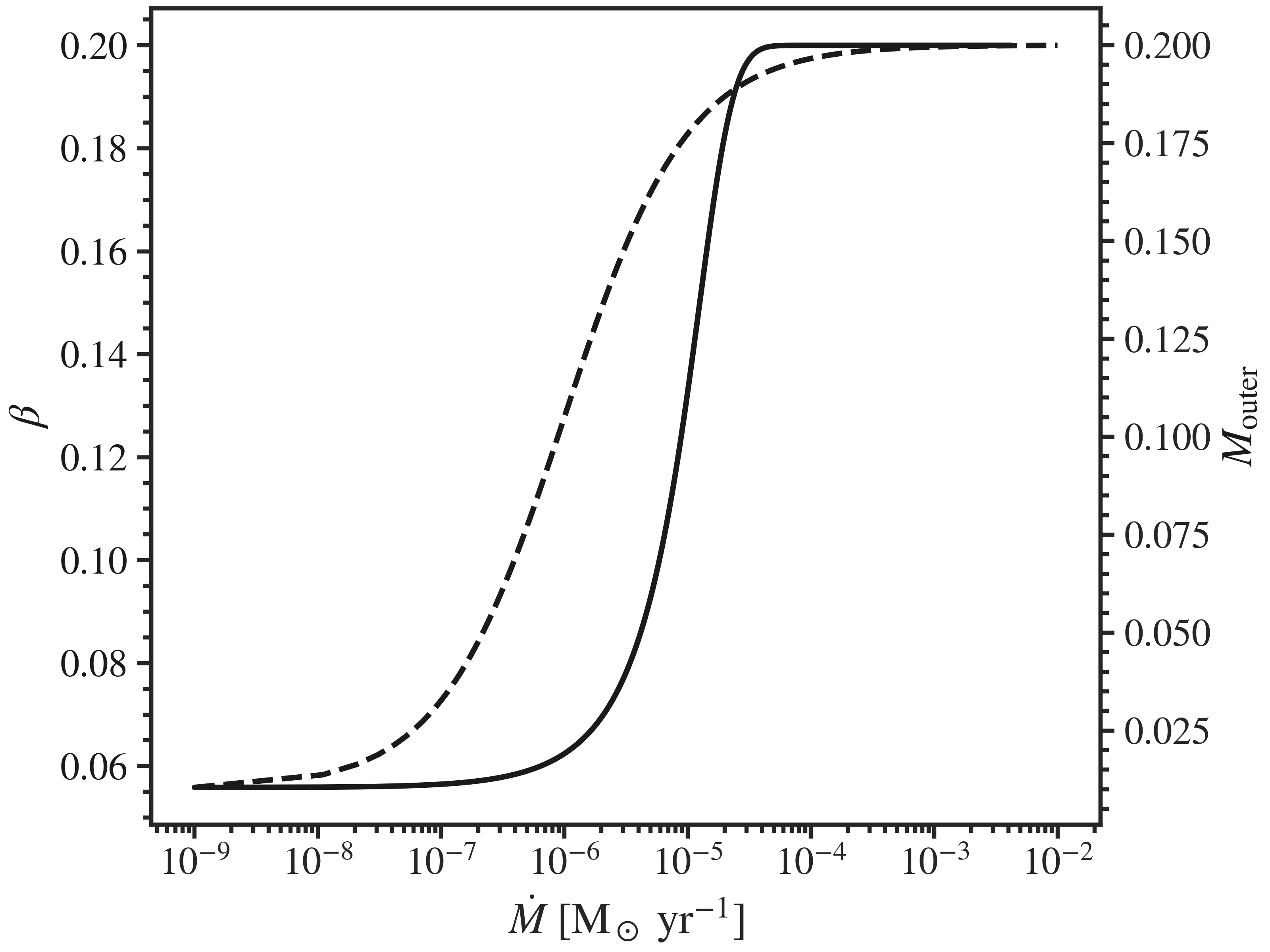}
    \caption{Thermal efficiency parameter $\beta$ (full line) (equation \ref{eq:beta}) and heat injection depth $M_\text{outer}$ (dashed line) (equation \ref{eq:Mouter}) against mass accretion rate $\dot{M}$.}
    \label{fig:app:beta_mouter}
\end{figure}

\begin{figure}
    \centering
    \includegraphics[width=1.0\linewidth]{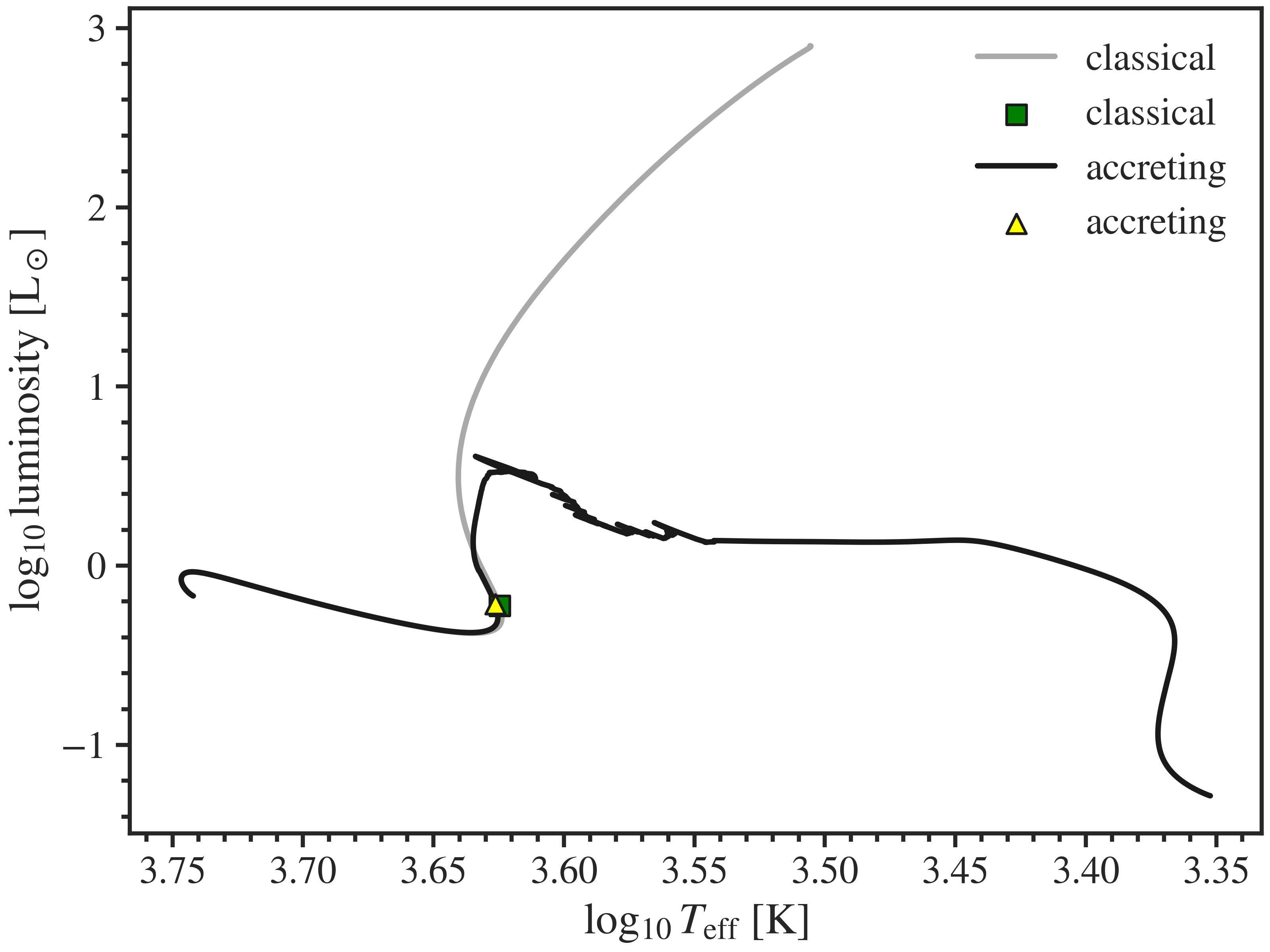}
    \caption{HR-diagram displaying the locations of the models used for which we compare propagation diagrams in Figure \ref{fig:prop}.}
    \label{fig:HR_classical_vs_accreting_prop}
\end{figure}


\bibliography{sample701}{}
\bibliographystyle{aasjournalv7}



\end{document}